\documentclass[prb,twocolumn,floatfix,epsf,psfig]{revtex4}
\pdfoutput=1 
\usepackage{epsf}
\usepackage{psfig}
\usepackage{graphicx}
\renewcommand{\figurename}{Figure}

\begin{document}
\title{Influence of anti-site disorder and electron-electron correlations on  the electronic structure of CeMnNi$_4$ }

	\author{Pampa Sadhukhan$^{1,\dagger}$,  Sunil Wilfred D$^{\prime}$Souza$^{1,{\dagger}*}$, Vipin Kumar Singh$^{1}$,  Rajendra Singh Dhaka$^{1a}$, Andrei Gloskovskii$^{2}$,  Sudesh Kumar Dhar$^{3}$, Pratap Raychaudhuri$^{3}$,  Ashish Chainani$^{4}$, Aparna Chakrabarti$^{5,6}$, Sudipta Roy Barman$^{1}$}
	\affiliation{$^{1}$UGC-DAE Consortium for Scientific Research, Khandwa Road, Indore 452001, Madhya Pradesh, India}
	\affiliation{$^{2}$Deutsches Elektronen-Synchrotron DESY, Notkestrasse 85, D-22607 Hamburg, Germany}
	\affiliation{$^{3}$Department of Condensed Matter Physics and Materials Science, Tata Institute of Fundamental Research, Homi Bhabha Road, Colaba, Mumbai 400005, India}
				\affiliation{$^{4}$National Synchrotron Radiation Research Center, Hsinchu 30076, Taiwan	 }
	\affiliation{$^{5}$Theory and Simulations Laboratory,  Raja Ramanna Centre for Advanced Technology, Indore, 452013, India}
		\affiliation{$^{6}$Homi Bhabha National Institute, Anushakti Nagar, Mumbai,  400094, India}
			
			\affiliation{$^{\dagger}$The authors have  contributed equally to this work.}

	\begin{abstract}
CeMnNi$_4$ exhibits an unusually large spin polarization, but its origin has baffled researchers for more than a decade. We use bulk sensitive hard x-ray photoelectron spectroscopy (HAXPES) and density functional theory  based on the Green's function technique to demonstrate the importance of electron-electron correlations  of both the Ni 3$d$  ($U_{Ni}$) and Mn 3$d$ ($U_{Mn}$) electrons in explaining  the valence band of this multiply correlated material. We show that  Mn-Ni anti-site disorder as well as $U_{Ni}$ play crucial role in enhancing its spin polarization:  anti-site disorder broadens a Ni 3$d$ minority-spin peak close to the Fermi level ($E_F$), while an increase in $U_{Ni}$ shifts it towards $E_F$, both leading to a significant increase of minority-spin states at $E_F$. Furthermore, rare occurrence of  a valence state transition between the bulk and the surface is demonstrated highlighting the importance of HAXPES  in resolving the electronic structure of materials unhindered by surface effects.

	\end{abstract}	

		\maketitle

In recent years, hard x-ray photoelectron spectroscopy (HAXPES) has turned out to be a reliable tool to study the electronic structure of correlated systems, thin films and  buried interfaces of  materials, thus providing new insights into their physical  properties\cite{Fadley10,Wocik16,Grayandothers}. In this work, we present the first study of the electronic structure of CeMnNi$_4$, an interesting  material with 
~large spin transport polarization of 66\%\cite{Singh06}, using HAXPES and  density functional theory calculations based on the spin polarized relativistic Korringa-Kohn-Rostoker (SPRKKR) method\cite{Ebert}. CeMnNi$_4$ has a cubic MgCu$_4$Sn-type structure\cite{Dhiman07}; it is ferromagnetic with a  magnetic moment of 4.95$\mu_B$ and Curie temperature of 140~K\cite{Singh06}.  These encouraging properties of CeMnNi$_4$  started a flurry of activity aimed at understanding its electronic structure\cite{Mazin06Voloshina06,Lahiri10,Bahramy10}.  However, no  photoemission study of its electronic structure has been reported to date, and the theoretical studies so far have been unable to explain the different aspects of its electronic structure and its spin polarization  in particular.  The early density functional theory (DFT) calculations\cite{Mazin06Voloshina06}  reported a spin polarization\cite{p0formula} ($P_0$) value of  about  16-20$\%$; and the  much larger experimental polarization was attributed to  disorder or non-stoichiometry of the specimens.  In fact, in a subsequent x-ray absorption fine structure  (XAFS) study, about  6$\%$ Mn-Ni anti-site disorder was reported\cite{Lahiri10}.   The authors also performed a DFT calculation using the pseudopotential method as implemented in the VASP code including an ordered anti-site defect configuration of nearest neighbour Ni and Mn that were site-exchanged. Thus, in this approach, the effect of randomly disordered anti-site defects is not taken into account. 
~Their results however showed a significant increase in $P_0$, which was not related to disorder,  but  rather to enhanced minority spin states of the site-exchanged Mn 3$d$ partial density of states (PDOS) due to hybridization with neighboring Ni atom\cite{Lahiri10}. On the other hand,  another DFT calculation that considered electron-electron correlation of the Mn 3$d$ electrons ($U_{Mn}$) but no anti-site defect showed that $P_0$ increases with $U_{Mn}$\cite{Bahramy10}. In the absence of any photoemission study  and its direct comparison with theory that addresses the influence of both anti-site disorder and correlation, their role in determining the electronic structure and spin polarization of  CeMnNi$_4$ has remained an unresolved question until date. 

In this letter, we show that both anti-site disorder and electron-electron correlations for Ni 3$d$ ($U_{Ni}$) and Mn 3$d$ ($U_{Mn}$) electrons have a crucial influence on the bulk electronic structure of CeMnNi$_4$. In addition, since $U_{Ce}$ is typically taken to be about 7 eV in Ce intermetallics\cite{Imer87}, CeMnNi$_4$ can be regarded as a multiply correlated system, further complicated by the presence of inherent disorder\cite{Lahiri10}. 
 $U_{Ni}$ and $U_{Mn}$   are responsible for determining the energy positions of the peaks in the valence band (VB) and their optimum values ($U_{Mn}$= 4.5 eV, $U_{Ni}$= 6.5 eV) are obtained by the best agreement between theoretically calculated and the experimental HAXPES VB. A surprising result is that the large $P_0$ of CeMnNi$_4$ has two origins: the  anti-site disorder ($x$) and $U_{Ni}$. The former broadens  a minority spin Ni 3$d$ peak close to $E_F$, while  the latter shifts it towards $E_F$. Thus, in both cases, the minority spin total DOS at $E_F$ ($n_\downarrow$($E_F$))  increases, while the majority spin total DOS ($n_\uparrow$($E_F$)) remains essentially unchanged, resulting in a clear  enhancement of $P_0$. The total magnetic moment exhibits contrasting variation: a decrease with $x$ and an increase with $U_{Ni}$. 
Furthermore, rare occurrence of a valence state transition on the surface of a ternary material is demonstrated: a bulk mixed valent state transforms to a nearly trivalent Ce$^{3+}$ state  due to the weakened hybridization on the surface.\\ 
~~\\
\noindent{\it \bf Experimental and computational methods:} HAXPES  measurements were performed at the P09 beamline in PETRA III synchrotron center, Germany  on polycrystalline CeMnNi$_4$  ingot that was cleaved under ultra high vacuum  at 2$\times$10$^{-8}$ mbar pressure to expose a fresh surface. The spectra were recorded by using  Phoibos 225 analyzer 
~with 30 eV pass energy  at 50~K\cite{Andrei}. Photons were incident on the sample at a grazing angle (10$^{\circ}$) and the photoelectrons were collected in the nearly normal emission geometry. The total instrumental resolution (including both source and analyzer contributions),  obtained from the least square fitting of the Au Fermi edge  in electrical contact with the specimen, is 0.26 eV. 
~CeMnNi$_{4}$ ingot was  prepared by an arc melting method and characterized for its structure using x-ray diffraction,  as discussed in Ref.~\onlinecite{Singh06}.

The bulk ground state properties of  CeMnNi$_4$  have been calculated in $F$$\bar{4}$$3m$ symmetry using the experimental lattice parameter ($a$= 6.9706\AA) as determined by neutron powder diffraction at 17 K\cite{Dhiman07}. Disordered Mn-Ni anti-site defects have been considered  by setting the 16$e$ site occupations to 1-0.25$x$ for Ni$_{\rm Ni}$ and 0.25$x$ for Mn$_{\rm Ni}$, while the occupancies at the 4$c$ site were set to $1-x$ for Mn$_{\rm Mn}$ and $x$ for Ni$_{\rm Mn}$, where X$_{\rm Z}$ refers to a X atom at a Z  atom site (X, Z= Ni, Mn).  Here, $x$ quantifies the amount of anti-site disorder as the fraction of Mn atoms occupying the Ni sites. In this work, we have  varied $x$ from 0 to 0.12. The structures are shown in Fig.~S1 of SM. 

Self-consistent band structure calculations were carried out using  fully relativistic SPRKKR method in the atomic sphere approximation\cite{Ebert}. The exchange and correlation effects were incorporated within the generalized gradient approximation framework.\cite{Perdew96}  The electron-electron correlation has been taken into account as described in the LSDA+U scheme\cite{Ebert03}. The parameters of screened on-site Coulomb interaction  $U$  for all the components ($U_{Ni}$, $U_{Mn}$ and $U_{Ce}$)  have been  varied  up to  7 eV, with the exchange interaction $J$ fixed at 0.8 eV. The static double counting of LSDA+U approach has been corrected using the atomic limit scheme. The angular momentum expansion up to $l_{max}$= 4 has been used for each atom. The energy convergence criterion and coherent potential approximation  tolerance has been set to 10$^{-5}$ Ry. Brillouin zone integrations were performed on a 36$\times$36$\times$36 mesh of $k$-points in the irreducible wedge of the Brillouin zone.  We have employed Lloyd's formula, which provides an accurate determination of the Fermi level and density of states\cite{Llyod}. For calculating the angle integrated VB spectrum, all the PDOS contributions from $\textit{s}$, $\textit{p}$, $\textit{d}$ and $\textit{f}$ states  of Ce, Mn and Ni  were  multiplied with their corresponding photoemission cross-sections\cite{Yeh85} and then added. 
~This is multiplied by the Fermi function and convoluted with the instrumental resolution  and an energy dependent lifetime broadening  0.01$\times$($E_B$-$E_F$)\cite{Barman95} to obtain the VB.

\begin{figure}[tb]
	\includegraphics[width=98mm,keepaspectratio]{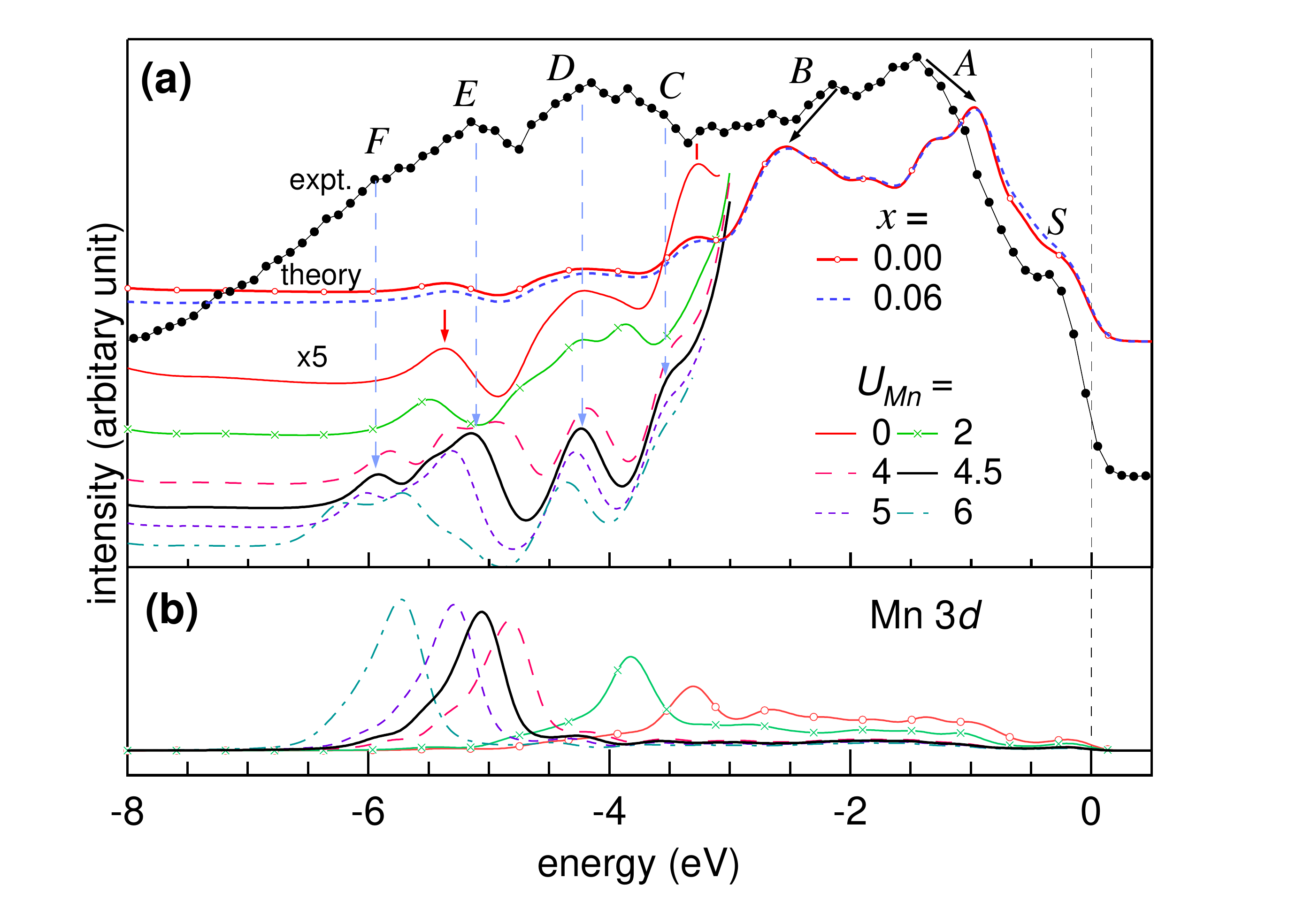} 
		
	\caption{(a) The  valence band (VB) HAXPES spectra of CeMnNi$_4$ at 50 K using 8 keV photon energy (black filled circles) compared with the calculated VB spectra for  $x$= 0 (no disorder) and $x$= 0.06 (6\% Mn-Ni anti-site disorder). The VB's calculated with different $U_{Mn}$ are shown in the -3 to -8 eV range, where $x$= 0, $U_{Ni}$= $U_{Ce}$= 0 eV. The spectra are staggered along the vertical axis, zero of the horizontal scale corresponds to the Fermi level ($E_F$).   (b)  Mn 3$d$ contribution to the calculated VB  as a function of $U_{Mn}$.}
	\label{mnu}
\end{figure}
~~\\
 \noindent{\it \bf Valence band of CeMnNi$_4$:}
 The   VB spectrum  recorded with 8 keV photon energy at 50 K shows a step ($S$) close to $E_F$ at -0.4 eV; peaks at -1.5 ($A$),  -2.2 ($B$), -3.6 ($C$), -4.2 ($D$), -5.2~eV ($E$)  and  a weak shoulder at -6 eV ($F$) (Fig.~\ref{mnu}(a)). In order to ascertain their origin and study the influence  of disorder on the spectral shape,  we have calculated the VB spectra without (red line with open circles, $x$= 0) and with 6\% Mn-Ni anti-site disorder (blue dashed line, $x$= 0.06). 6\% disorder is considered because a previous XAFS study\cite{Lahiri10} inferred a disorder of this magnitude on a specimen that was prepared by the same procedure as ours.
  As discussed above, 
  ~the VB has been calculated from  the partial DOS (PDOS)  in Fig.~S2.  We find that disorder results in a small but finite broadening of the VB, but it has no effect on the position of the peaks.   Comparison of the calculated VB with  HAXPES  shows glaring differences: the peaks corresponding to $A$ and $B$ (black arrows) are positioned at higher and lower energies, respectively and thus their separation  (1.6 eV) is significantly larger compared to experiment  (0.7 eV). The peak at -5.4 eV (red arrow) is shifted  $w.r.t.$ peak $E$ of the experimental VB, the peak at  -3.3 eV (red tick) appears at a dip, while  there is no peak in the theory corresponding to $F$ (see the blue dashed arrows). In Fig.~S2, DOS  calculated with disorder up to $x$= 0.12 ($i.e.$ 12\% aniti-site disorder) show increased broadening, but the positions of all the peaks  remain unchanged. 
  
Thus, it is obvious from the above discussion that disorder is unable to explain the VB. So, we examine the possible role of correlation  starting with $U_{Mn}$.  As $U_{Mn}$ is increased, interesting modifications in  the -3 to -6 eV region is  observed in Fig.~\ref{mnu}(a), which are primarily related to the systematic changes in the Mn 3$d$ PDOS (Fig.~\ref{mnu}(b) and PDOS in Fig.~S3).  At $U_{Mn}$= 0, the Mn 3$d$ states are delocalized over 0 to -5 eV with the most intense peak at -3.3 eV.  Increase of $U_{Mn}$  narrows the Mn 3$d$ PDOS, the peak intensity increases  and it shifts by a large amount to lower energies {\it i.e.} away from $E_F$ ($e.g.$ -5.2 eV for $U_{Mn}$= 4.5 eV). The best agreement with experiment in the -3 to -6 eV region is obtained for  $U_{Mn}$= 4.5 eV (black line), where the peaks at -3.6, -4.2, -5.2 and  -6 eV appear at the same positions as $C$, $D$, $E$, and $F$,  respectively of the experimental VB, as shown by the blue dashed arrows in Fig.~\ref{mnu}(a).   
~ The Mn 3$d$ states contribute primarily to  the peak $E$, however, its intensity is relatively less due to  smaller photoemission cross-section of Mn 3$d$ with respect to Ni 3$d$  at 8 keV\cite{Yeh85}.

\begin{figure}[tb]
	\includegraphics[width=98mm,keepaspectratio]{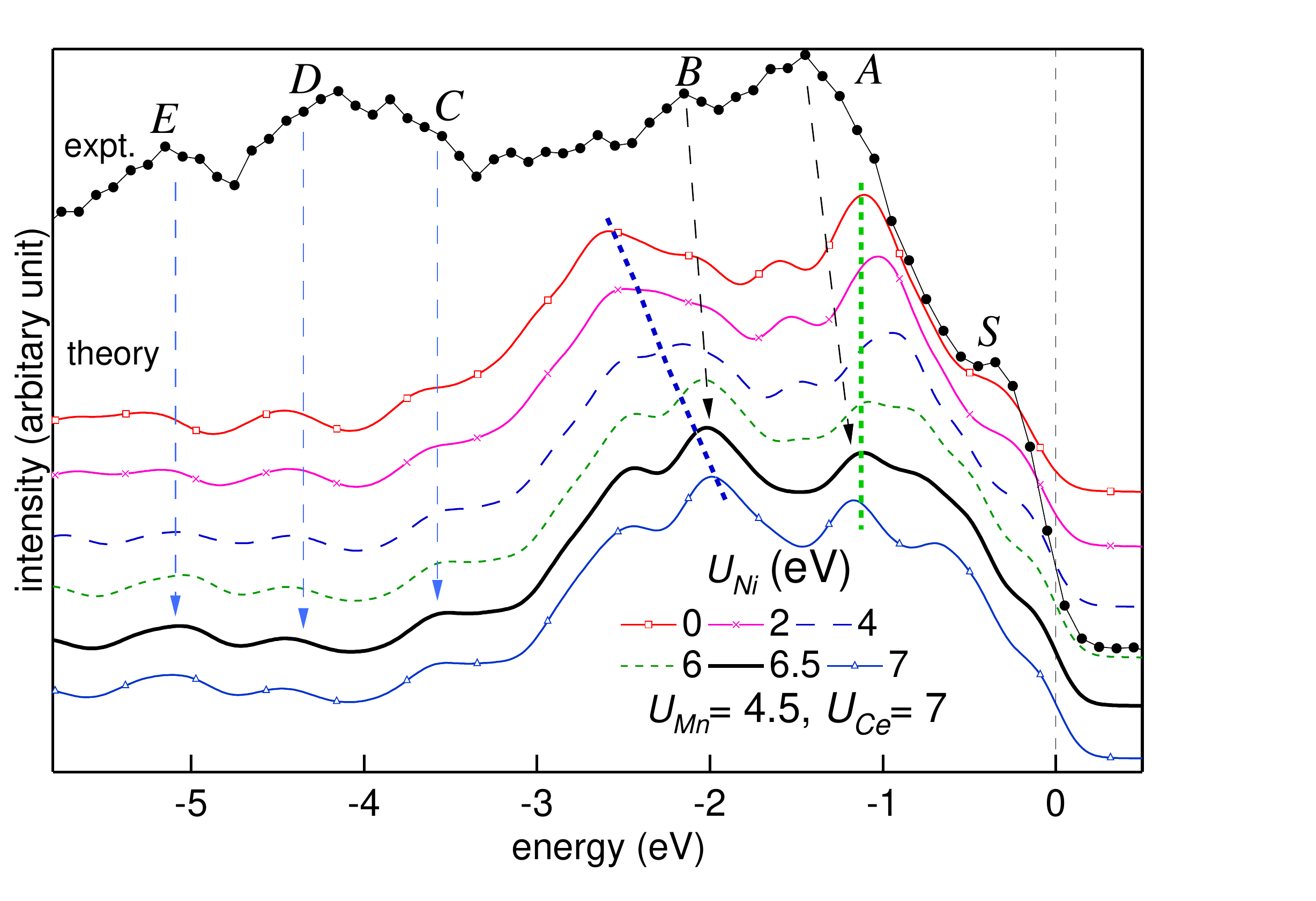}
	
	\caption{The  valence band HAXPES spectrum of Fig.~1 (black filled circles) 
		~ compared with calculated VB spectra as a function of  $U_{Ni}$, with  $U_{Mn}$= 4.5 eV, $U_{Ce}$= 7 eV and $x$= 0.} 
	\label{niu}
\end{figure}

Although $U_{Mn}$= 4.5 eV provides a good agreement for peaks $C$-$F$, the positions of the  peaks $A$ and $B$ are not well reproduced, and these  remain unaltered with $U_{Mn}$ (Fig.~S3). It is evident that $A$ and $B$ originate primarily from Ni 3$d$ states, and so we calculate the VB by introducing $U_{Ni}$, with $U_{Mn}$ fixed at 4.5 eV.  We find that as $U_{Ni}$ increases, the peak at -2.6 eV shifts to higher energy {\it i.e.} towards $E_F$ (blue dashed line) and  appears close to the position of peak $B$ for $U_{Ni}$= 6.5 eV (Fig.~\ref{niu}, see Fig.~S4 for PDOS). On the other hand, the peak at -1.1 eV initially shifts to higher energies and eventually shifts back to lower energy (green dashed line) towards peak $A$. The separation of these two peaks is lowest (0.8 eV)  at  $U_{Ni}$= 7 eV.  However, for $U_{Ni}$= 7 eV, a new peak appears at -0.7 eV in disagreement with experiment. Thus, we conclude that the best agreement is observed for  $U_{Ni}$= 6.5 eV, where the positions as well as the separation (0.9 eV) of the calculated peaks agree well with $A$ and $B$ (black dashed arrows in Fig.~\ref{niu}). Note that the peaks in the -3 to -6 eV region are hardly affected by $U_{Ni}$.  

It is to be noted that in Fig.~\ref{niu} we  also consider a value of $U_{Ce}$ (= 7 eV) for the Ce 4$f$ electrons that is generally observed in Ce intermetallic compounds\cite{Imer87}. However, $U_{Ce}$ does not have any discernible effect on the occupied states and the VB, since the  Ce 4$f$ peak appears mostly above $E_F$ at 0.9 eV  for $U_{Ce}$= 0 
~(Fig.~S3(c))
and shifts to higher energy (1.2 eV) for $U_{Ce}$= 7 eV 
~(Fig.~S4(a)).  Thus,   due to the significant variation of Ni and Mn 3$d$ states with $U_{Ni}$ and $U_{Mn}$, respectively and taking $U_{Ce}$ from literature\cite{Imer87},  we are able to determine the optimum values of $U$ for CeMnNi$_4$ to be: $U_{Mn}$= 4.5 eV, $U_{Ni}$= 6.5 eV and $U_{Ce}$= 7 eV  (referred henceforth as $U(4.5,6.5,7)$). The partial contributions of the different PDOS to each of the peaks in the VB for $U(4.5,6.5,7)$ are shown in Fig.~S5 of SM. 

\begin{figure}[tb]
	\includegraphics[width=93mm,keepaspectratio]{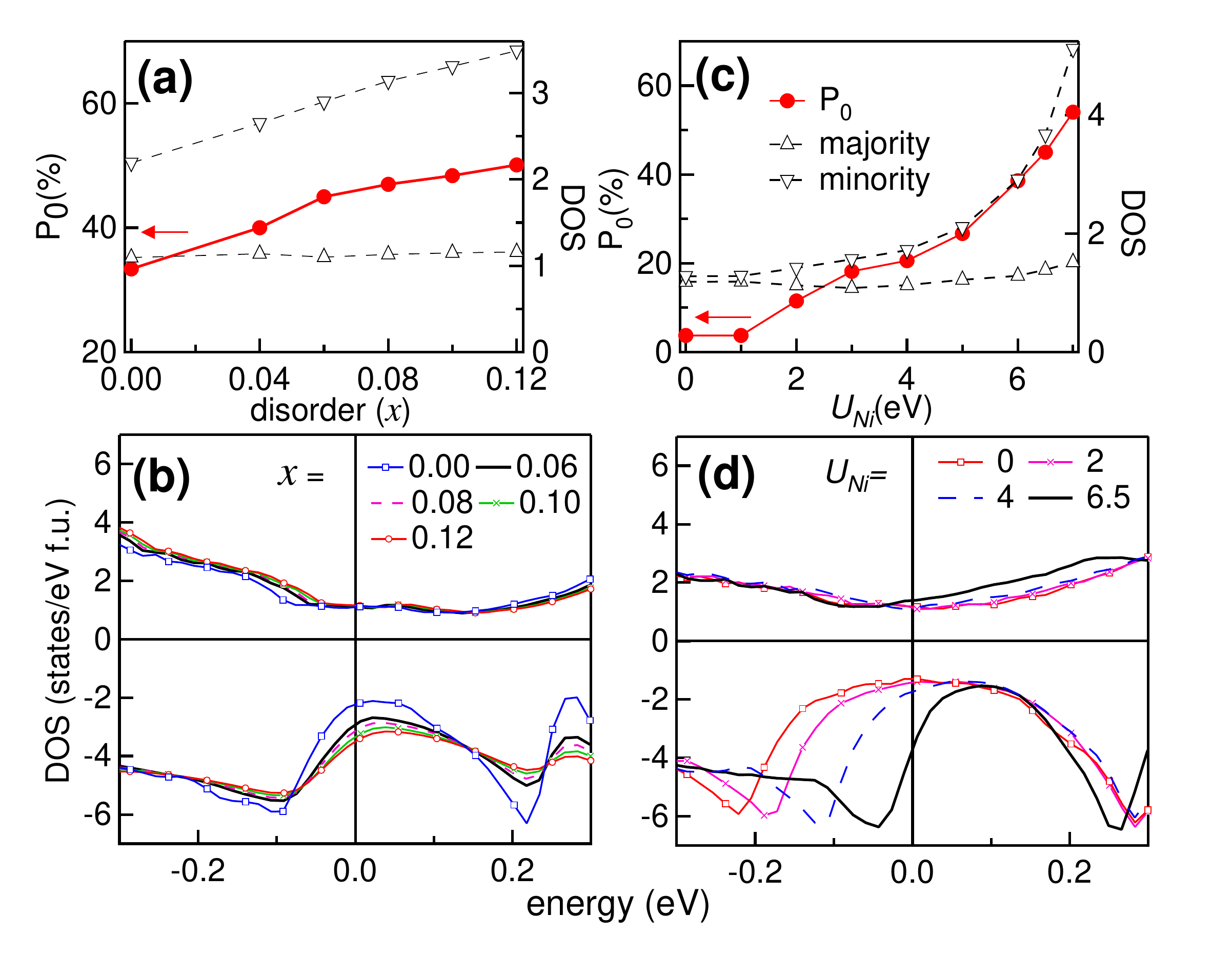}
	\caption{Spin polarization  ($P_0$), majority ($n_\uparrow$($E_F$)) and minority ($n_\downarrow$($E_F$)) spin total DOS at $E_F$   (a) as a function of disorder ($x$) with  $U_{Ni}$= $U_{Mn}$= $U_{Ce}$= 0; and (c) as a function of $U_{Ni}$, where  $U_{Mn}$= 4.5 eV, $U_{Ce}$= 7 eV and $x$= 0.
		Majority
		~and minority 
		~spin total DOS  around  $E_F$  corresponding to (a) and (c) as a function of (b) $x$ and (d) $U_{Ni}$, respectively.}
	\label{spol}
\end{figure}
 ~~\\
 ~~\\
 \noindent{\it \bf  Spin polarization and magnetic moments:}
 We find that the Mn-Ni anti-site disorder has an unexpected positive effect on the spin polarization ($P_0$).  As shown in Fig.~\ref{spol}(a) and Table\,{\bf I} of SM, $P_0$ exhibits a  monotonic increase with  $x$, reaching a value of 45\% (50\%) for $x$= 0.06 (0.12). This is an important result since  in half metals and Heusler alloys, a low  experimental value of $P_0$  is generally attributed to disorder\cite{hm}.  
 In order to understand the reason for this unusual behavior,  we show the spin polarized total DOS around $E_F$ in Fig.~\ref{spol}(b). A peak in the minority spin DOS close to $E_F$ at -0.1~eV  progressively broadens and also shifts by a small amount ($\approx$ 15 meV) towards $E_F$ resulting in increase of $n_\downarrow$($E_F$) with $x$. On the contrary, the structureless majority spin  DOS and consequently $n_\uparrow$($E_F$) remain almost unchanged. Thus, this contrasting behavior of  $n_\downarrow$($E_F$) and $n_\uparrow$($E_F$)  brings about the increase of $P_0$ with $x$  (Fig.~\ref{spol}(a)). Table\,{\bf I} of SM defines and shows the partial contributions from Ni 3$d$ ($P_{0_{{\rm Ni}3d}}$), Mn 3$d$ ($P_{0_{{\rm Mn}3d}}$) and Ce 4$f$ ($P_{0_{{\rm Ce}4f}}$) PDOS to $P_0$ for different $x$, and we find that $P_0$ increases solely because of $P_{0_{{\rm Ni}3d}}$. This is also confirmed in Fig.~S6 where  the peak in the minority spin DOS  is  clearly dominated by Ni 3$d$ PDOS (black tick).

Turning to the influence of $U$ on $P_0$  (Fig.~\ref{spol}(c)), we find that it  increases   with $U_{Ni}$ from about 3.8\% for $U$(4.5,0,7)  to  45\%  for $U_{Ni}$= 6.5 eV $i.e$ for the optimum $U$(4.5,6.5,7). This is related to  increase of $n_\downarrow$($E_F$)  due to  a significant shift of the  minority spin total DOS peak  towards $E_F$  from -0.2 to -0.05 eV (Fig.~\ref{spol}(d)). Clearly the total DOS is dominated by Ni 3$d$, black ticks in Fig.~S7 show how the minority spin Ni 3$d$ PDOS peak shifts with $U_{Ni}$. In contrast, the majority spin total DOS is structureless and $n_\uparrow$($E_F$) remains almost unchanged (Fig.~\ref{spol}(c,d)). The partial contributions to $P_0$ for different $U_{Ni}$ clearly show that the  increase in $P_0$ is entirely due to $P_{0_{{\rm Ni}3d}}$ (Table\,{\bf I} of SM).

 Due to disorder, the Ni 3$d$ minority spin peak 
 ~will  broaden and also possibly shift by small amount towards $E_F$ and thus significantly increase  $n_\downarrow$($E_F$) because of its proximity to $E_F$ ($e.g.$ at -0.05 eV for $U$(4.5,6.5,7)). On the other hand,  $n_\uparrow$($E_F$) would remain unchanged due to the nearly  flat nature of the majority spin total DOS. Thus, disorder would further increase $P_0$, and assuming that its  effect is independent of $U$,   we  estimate  $P_0$ for $U$(4.5,6.5,7) to increase from 45\% to $>$55\% ($>$60\%)  for $x$= 0.06 (0.12). This is in good agreement with the  experimental value of 66\%,  given the fact that the measurements were performed  in the diffusive limit\cite{Singh06} and here we calculate the static spin polarization. 

\begin{figure}[tb]
	    \includegraphics[width=93mm,keepaspectratio]{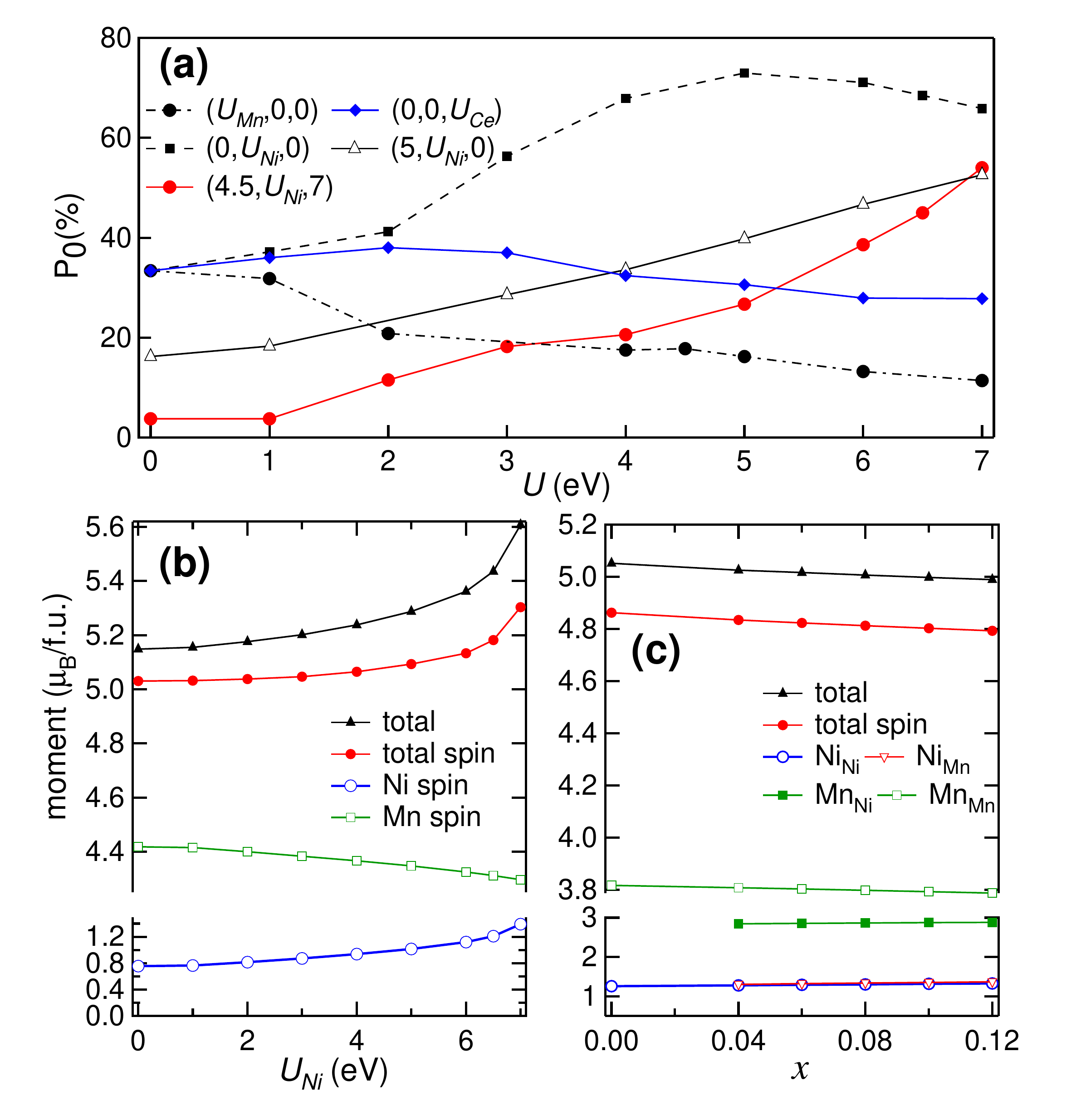}
		\caption{(a) Spin polarization P$_{0}$ as a function of electron-electron correlation $U$ for Ni 3$d$  ($U_{Ni}$), Mn 3$d$  ($U_{Mn}$) and Ce 4$f$ electrons ($U_{Ce}$). $P_0$ is plotted as a function of $U$  shown as a triplet ($U_{Mn}$, $U_{Ni}$, $U_{Ce}$), where the fixed $U$ values in the triplet are indicated by numbers in eV. For example, (5, $U_{Ni}$,0) means $U_{Ni}$  varies from 0 to 7 eV with $U_{Mn}$ and  $U_{Ce}$ fixed at 5 eV and 0 eV, respectively. The  total (spin plus orbital) moment, the total spin only moment of CeMnNi$_4$  and the local spin magnetic moments of Mn and Ni are plotted  (b) as a function of $\textit{U}$$_{Ni}$ with $\textit{U}$$_{Mn}$= 4.5 eV, $\textit{U}$$_{Ce}$= 7 eV and $x$= 0; and (c) as a function of disorder ($\textit{x}$). 
			~In all cases, the Ce atom possess a small opposite moment of  -0.2 $\mu_B$.}
	\label{S8&9}
	\end{figure}

We have also studied how  $U_{Mn}$ and $U_{Ce}$ affects $P_0$ and find that both have detrimental effect: in Fig.~\ref{S8&9}(a), $P_0$($U_{Mn}$,0,0)  shows a decrease  from 33.4\% to 11.4\% with $U_{Mn}$ varying from 0 to 7 eV. In comparison, the effect of $U_{Ce}$ is milder with  $P_0$(0,0,$U_{Ce}$)  decreasing  from 33.4\% to 28\%. 
~ If  $U_{Mn}$ and $U_{Ce}$  are set to 0,   $P_0$ increases to a large value of 66\% for $U_{Ni}$= 7 eV $i.e.$ for $U(0,7,0)$ (black filled squares in Fig.~\ref{S8&9}(a)). On the other hand, a comparison of $P_0$($U_{Ni}$)  for  (0,$U_{Ni}$,0), (5,$U_{Ni}$,0), (4.5,$U_{Ni}$,7)  shows that  the extent of increase of $P_0$ is clearly arrested when $U_{Mn}$ and $U_{Ce}$ are non-zero. These results refute an earlier counterintuitive report\cite{Bahramy10}, which concluded that  $U_{Mn}$ increases $P_0$, while  neither $U_{Ni}$ nor $U_{Ce}$ have any influence on $P_0$ (see Supplementary discussion SD1).

The calculated  magnetic moments show that  the total moment of  CeMnNi$_4$ is quite large $e.g.$    5.43~$\mu_B$ for $U(4.5,6.5,7)$, the main contribution coming from the Mn spin moment (4.31~$\mu_B$).  Fig.~\ref{S8&9}(b)  shows that both the total moment as well as the Ni spin moment increase with $U_{Ni}$, $e.g$ for $U(4.5,0,7)$ the total moment (Ni spin moment)  is  5.15 (0.19) $\mu_B$,  whereas for $U(4.5,6.5,7)$ it is 5.43 (0.3) $\mu_B$.   The  increase in the Ni spin moment is because of  the shift of the Ni 3$d$ minority spin 
~states   towards $E_F$ (Fig.~\ref{spol}(d)) resulting in a decrease of the integrated occupied minority spin PDOS, while the majority spin PDOS remains largely unchanged.  It may be noted that the total moment of  5.43~$\mu_B$  for $U(4.5,6.5,7)$   is somewhat overestimated compared to the experimental value of  4.95~$\mu_B$ from magnetization measurement at 5~K\cite{Singh06}. 

Interestingly, we find that the  total magnetic moment decreases with increasing disorder  (Fig.~\ref{S8&9}(c)). This can be ascribed to the difference of the   Mn$_{\rm Ni}$ (Mn atom in Ni position)    and Mn$_{\rm Mn}$ (Mn atom in Mn position) 3$d$ spin-polarized PDOS, the latter having considerably reduced exchange splitting (Fig.~S2). This difference  is related to the change in hybridization due to different nearest neighbor configurations (Fig.~S1).   The local moment of Mn$_{\rm Ni}$ is thus substantially smaller (2.8~$\mu_B$) compared to Mn$_{\rm Mn}$ (3.8~$\mu_B$).  Although the local moments hardly vary, the proportion of Mn$_{\rm Ni}$ increases  with $x$, resulting in a decrease of the total moment. Thus, it can be argued that the overestimation  of the total moment by theory with  $U$(4.5,6.5,7) mentioned above   could be somewhat compensated  by its decrease caused by anti-site disorder. 

An additional interesting outcome of our study 
~is the demonstration of a valence state transition $i.e.$ a change of the valency of Ce between the bulk and the surface. Valence state transition could significantly alter the  surface electronic structure compared to the bulk. It was first  reported in Sm metal\cite{Wertheim78} 
~and later in binary Ce intermetallic compounds\cite{Laubschat90}. 
~ From the analysis of  the Ce 3$d$ core-level spectra using HAXPES and XPS and using a simplified version of the  Anderson single-impurity model\cite{Gunnarsson83} proposed by Imer and Wuilloud (IW)\cite{Imer87}, we show  that 
the Ce 4$f$ occupancy in the  ground state ($n_f$) turns out to be 0.8 in the bulk, indicating a mixed valent state with 20\% Ce in $f^0$ (Ce$^{4+}$) while 80\% in $f^1$ (Ce$^{3+}$) configuration, where  $f^{0}$ and $f^{1}$ are the satellite peaks in the Ce 3$d$ spectrum related to 3$d^{9}$4$f^{0}$ and 3$d^{9}$4$f^{1}$ final states, respectively\cite{Hillebrecht82,Fuggle83}. In contrast, from  the surface sensitive Ce 3$d$ XPS spectrum,  $n_f$ increases to 0.98 and thus the surface has predominantly  3$d^{9}$4$f^{1}$ (Ce$^{3+}$) ground state. Thus, in the bulk,  the Ce 4$f$ electron transfers to the valence states comprising  primarily of Ni 3$d$ states making CeMnNi$_4$ a  mixed valent system with 4$f$ occupancy  of $n_f$= 0.8. However, at the surface, the reduced hybridization between the  Ce 4$f$ and unsaturated 3$d$ states results in a lowering of the Ce 4$f$ states further below $E_F$. This increases the occupancy of the Ce 4$f$ level ($n_f$= 0.98) and results in the valence state transition. The detailed discussion on the valence state transition and comparison with surface sensitive XPS  is provided in the Supplementary discussion SD2.

 In conclusion, we settle the long standing debate about the electronic structure of CeMnNi$_4$. We establish the importance of both anti-site disorder and electron-electron correlation in explaining its intriguing properties. Our work  fundamentally alters the general notion that anti-site disorder is detrimental for spin polarization.     We hope it will motivate further experimental work on CeMnNi$_4$ and related materials, mainly because disorder could  be controlled and $P_0$ further enhanced. We find that the total magnetic moment exhibits contrasting behaviour, it decreases with $x$, but increases with $U_{Ni}$. 
 ~A valence state transition that originates due to the weakened hybridization on the surface is  demonstrated. Our study  highlights the power of HAXPES in combination with density functional theory for clarifying the electronic structure and properties of multiply-correlated materials with inherent anti-site disorder.


{\noindent\bf Acknowledgments:} The experiments were carried out at PETRA III of Deutsches Elektronen-Synchrotron, a member of Helmholtz-Gemeinschaft Deutscher Forschungszentren. Financial support by the Department of Science and Technology, Government of India within the framework of India@DESY collaboration is gratefully acknowledged. We would like to thank W. Drube and C. Narayana for support and encouragement.  S.W.D. gratefully acknowledges the financial support from CEDAMNF project (CZ.02.1.01/0.0/0.0/15-003/0000358), New Technologies Research Centre, University of West Bohemia, Czech Republic. A.C. thanks P.A. Naik, A. Banerjee  for support and encouragement and the Computer Centre of RRCAT, Indore  for providing the computational facility for a part of the work. 

\noindent {\it $^{*,a}$Present addresseses}: $^*$New Technologies Research Centre, University of West Bohemia, Univerzitn\'{\i} 8, CZ-306 14 Pilsen, Czech Republic; $^{a}$Department of Physics, Indian Institute of Technology Delhi, Hauz Khas, New Delhi 110016, India

\newpage
\setcounter{figure}{0}
\renewcommand{\figurename}{Fig.~S}
\begin{center} 
		{\it Supplementary material to the paper entitled:}\\
			~~\\
			
			{\bf Influence of anti-site disorder and electron-electron correlations on  the electronic structure of CeMnNi$_4$  }
~~\\
~~\\
~~\\
\noindent Pampa Sadhukhan$^{1,\dagger}$,  Sunil Wilfred D$^{\prime}$Souza$^{1,{\dagger}*}$, Vipin Kumar Singh$^{1}$,  Rajendra Singh Dhaka$^{1a}$, Andrei Gloskovskii$^{2}$,  Sudesh Kumar Dhar$^{3}$, Pratap Raichaudhuri$^{3}$,  Ashish Chainani$^{4}$, Aparna Chakrabarti$^{5}$, Sudipta Roy Barman$^{1}$\\
	$^{\dagger}$Both the authors have  contributed equally to this work.\\
\end{center}
~~\\
\noindent{ This Supplementary material contains seven figures (S1 to S7), two tables (TABLE-I and II) and two Supplementary discussions (SD1 and SD2, which include figures S8-S13 ).}

\begin{figure}[h]
	\begin{center}
		\includegraphics[width=85mm,keepaspectratio]{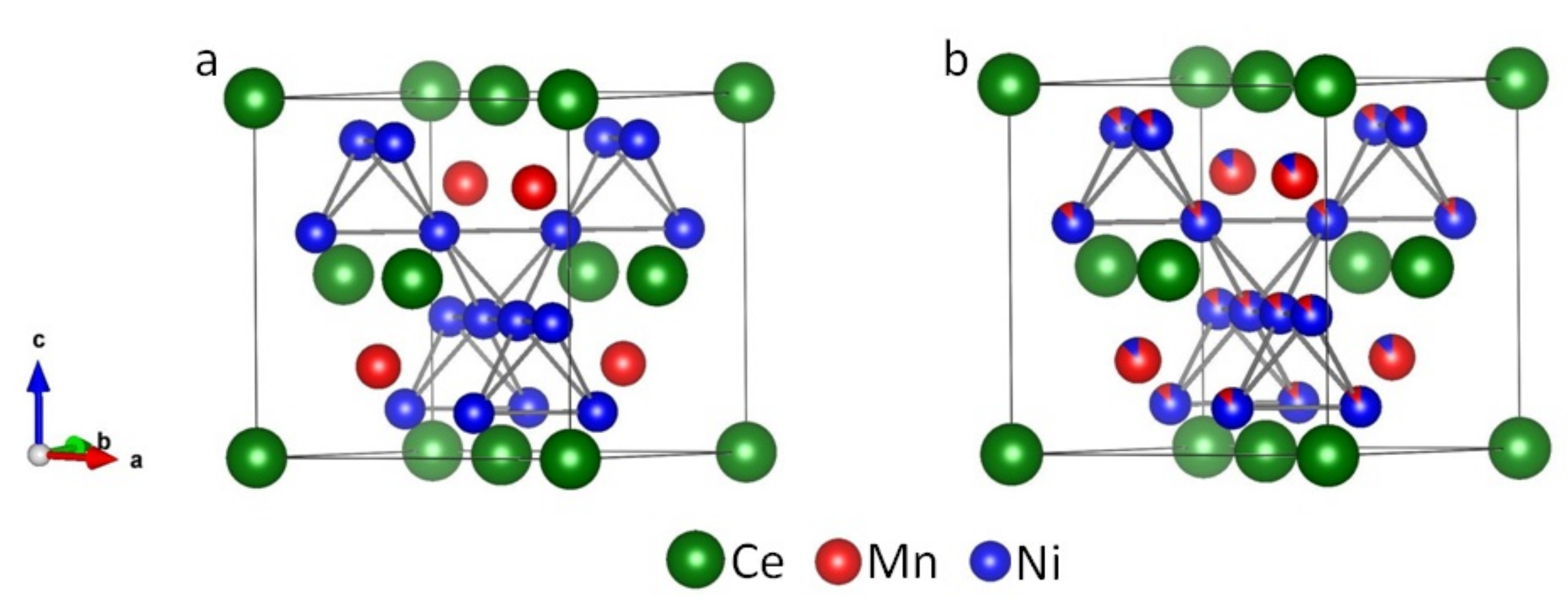}
		\caption{The crystal structure of  (a) ordered CeMnNi$_4$ with  Ce and Mn are placed at 4$a$ (0, 0, 0) and 4$c$ (0.25, 0.25, 0.25) sites, respectively, while Ni is placed at 16$e$ (0.624, 0.624, 0.624) site. The corresponding multiplicities of 4$a$, 4$c$ and 16$e$ atomic sites are  1, 1 and 4, respectively and  (b) Disordered CeMnNi$_4$ with 12\% ($x$= 0.12) Mn-Ni anti-site disorder. The crystallographic axes are represented by the arrows ($a, b, c$). The entire cubic structure has been rotated by 45$^{\circ}$ in the  clockwise direction to provide a better view. Note that the first nearest neighbor (nn) of Mn$_{\rm Mn}$ (Mn atom in Mn position)  are three Ni atoms, while the second nn are four Ce atoms. On the other hand, for Mn$_{\rm Ni}$ (Mn atom in Ni position), the first nn are six Ni atoms, while the second nn are three Mn$_{\rm Mn}$ atoms.}
	\end{center}
	\label{S2_structure}
\end{figure}

\begin{figure}[ht]
	\begin{center}
		\includegraphics[width=93mm,keepaspectratio]{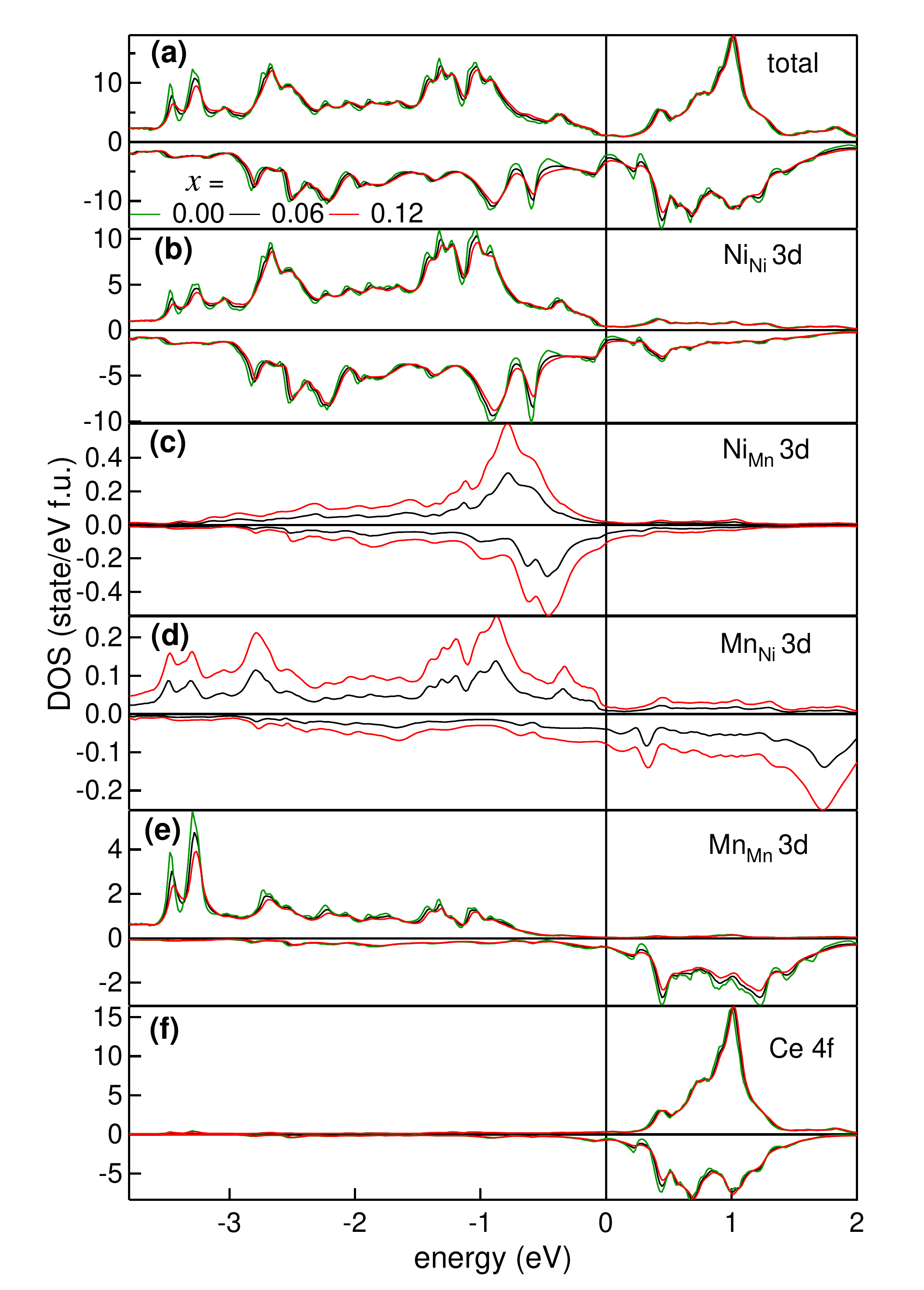}
		\caption{ Spin polarized (a) total DOS and the PDOS for (b) Ni$_{\rm Ni}$ 3$d$, (c) Ni$_{\rm Mn}$ 3$d$, (d) Mn$_{\rm Ni}$ 3$d$, (e) Mn$_{\rm Mn}$ 3$d$  and (f) Ce 4$f$ as a function of Mn-Ni anti-site disorder quantified by $x$= 0, 0.06 and 0.12.}
	\end{center}
	\label{S1}
\end{figure}

\begin{figure}[t]
	\begin{center}
		\includegraphics[width=93mm,keepaspectratio]{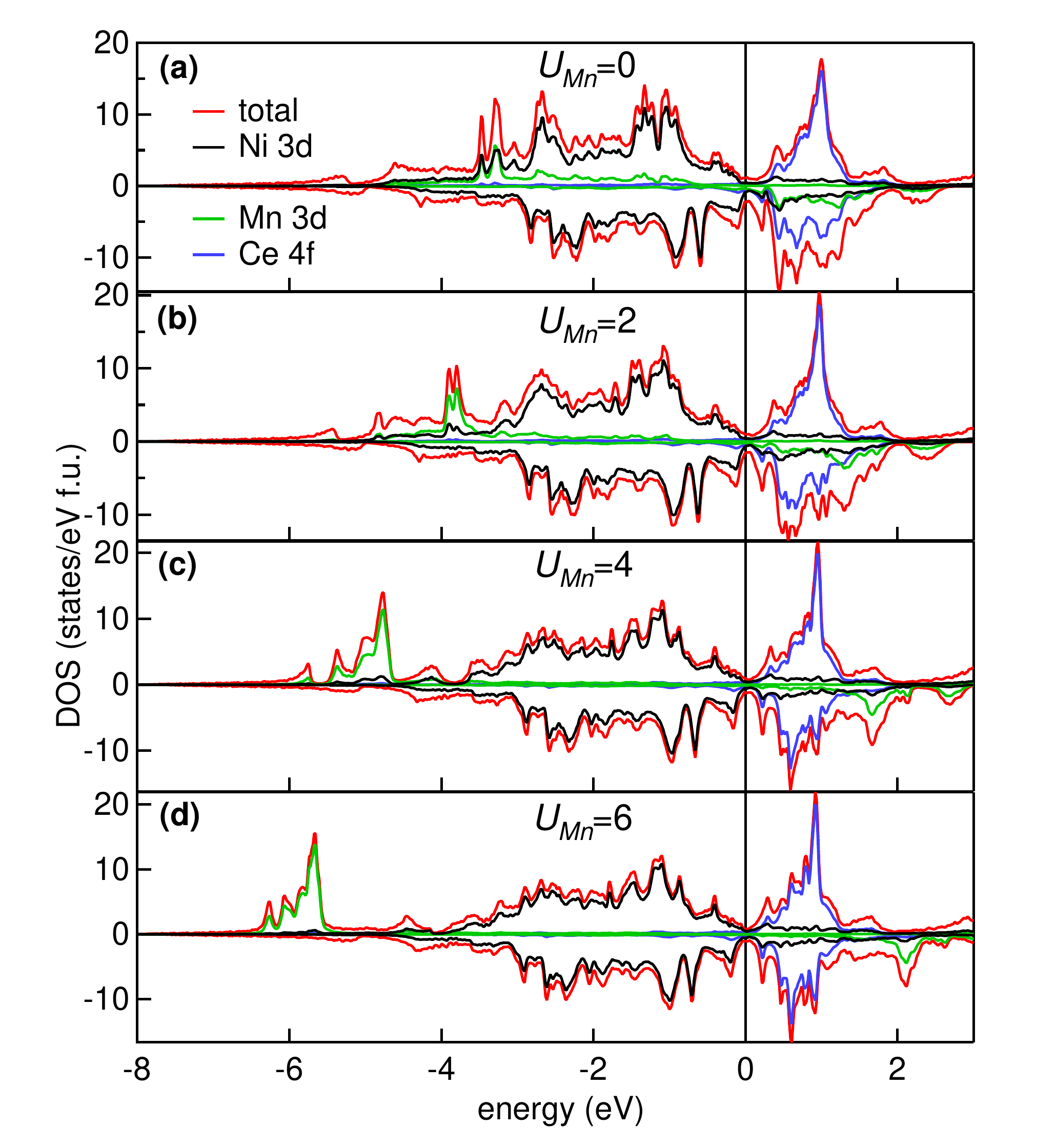} 
		
		\caption{ Spin polarized total DOS and Ni 3$d$, Mn 3$d$ and Ce 4$f$ PDOS as a function of $U_{Mn}$ with $U_{Ni}$= $U_{Ce}$= 0 where for (a) $U_{Mn}$= 0 eV, (b) $U_{Mn}$= 2 eV, (c) $U_{Mn}$= 4 eV, and (d) $U_{Mn}$= 6 eV.}
	\end{center}
	\label{S3}
\end{figure}

\begin{figure}[t]
	\begin{center}
		\includegraphics[width=93mm,keepaspectratio]{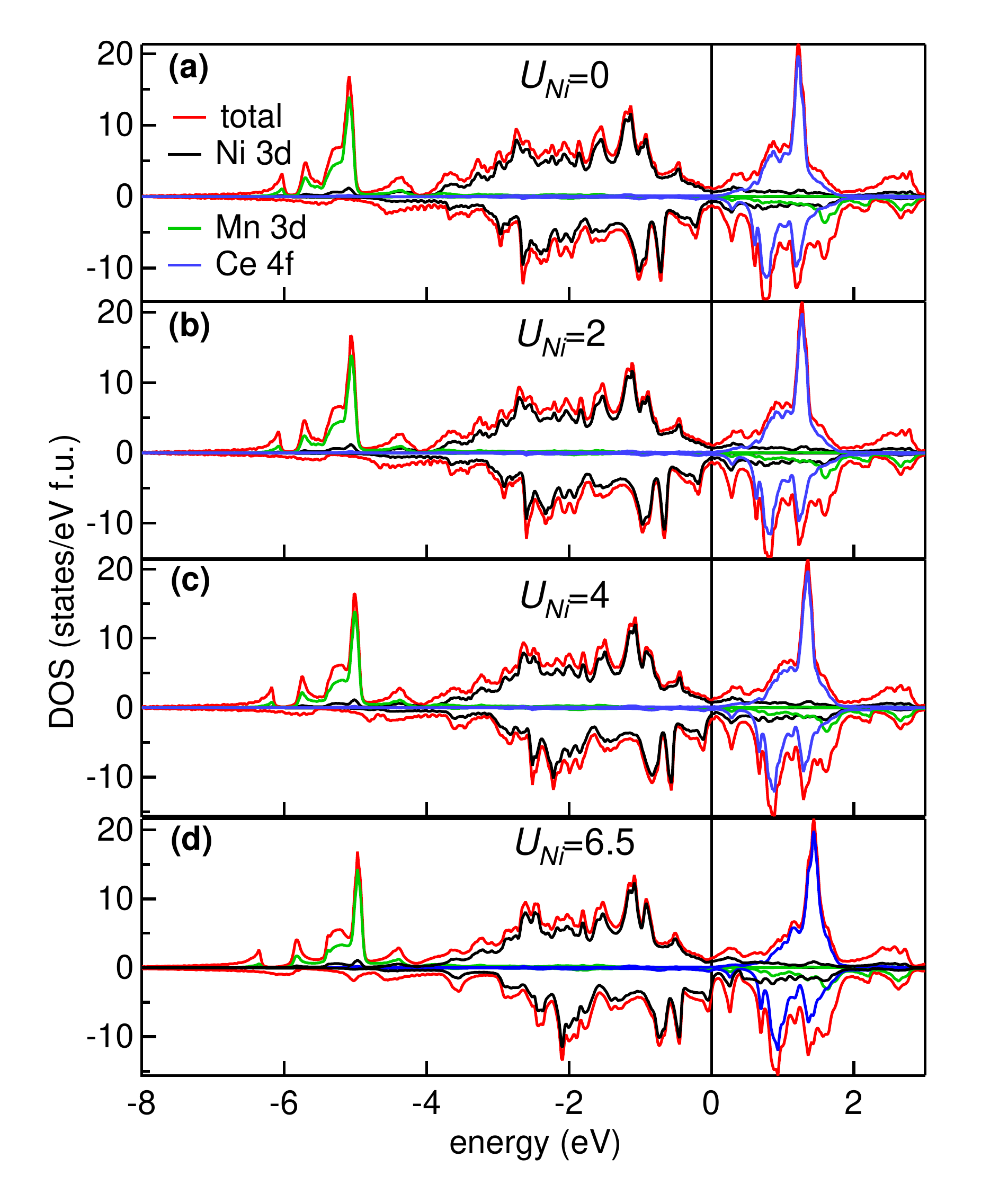} 
		\caption{Spin polarized total DOS and Ni 3$d$, Mn 3$d$ and Ce 4$f$ PDOS as a function of $U_{Ni}$ with fixed $U_{Mn}$= 4.5 eV and $U_{Ce}$= 7 eV where for (a) $U_{Ni}$= 0 eV, (b) $U_{Ni}$= 2 eV, (c) $U_{Ni}$= 4 eV and (d) $U_{Ni}$= 6.5 eV.}
	\end{center}
	\label{S4}
\end{figure}

\begin{figure}[t]
	\begin{center}
		\includegraphics[width=93mm,keepaspectratio]{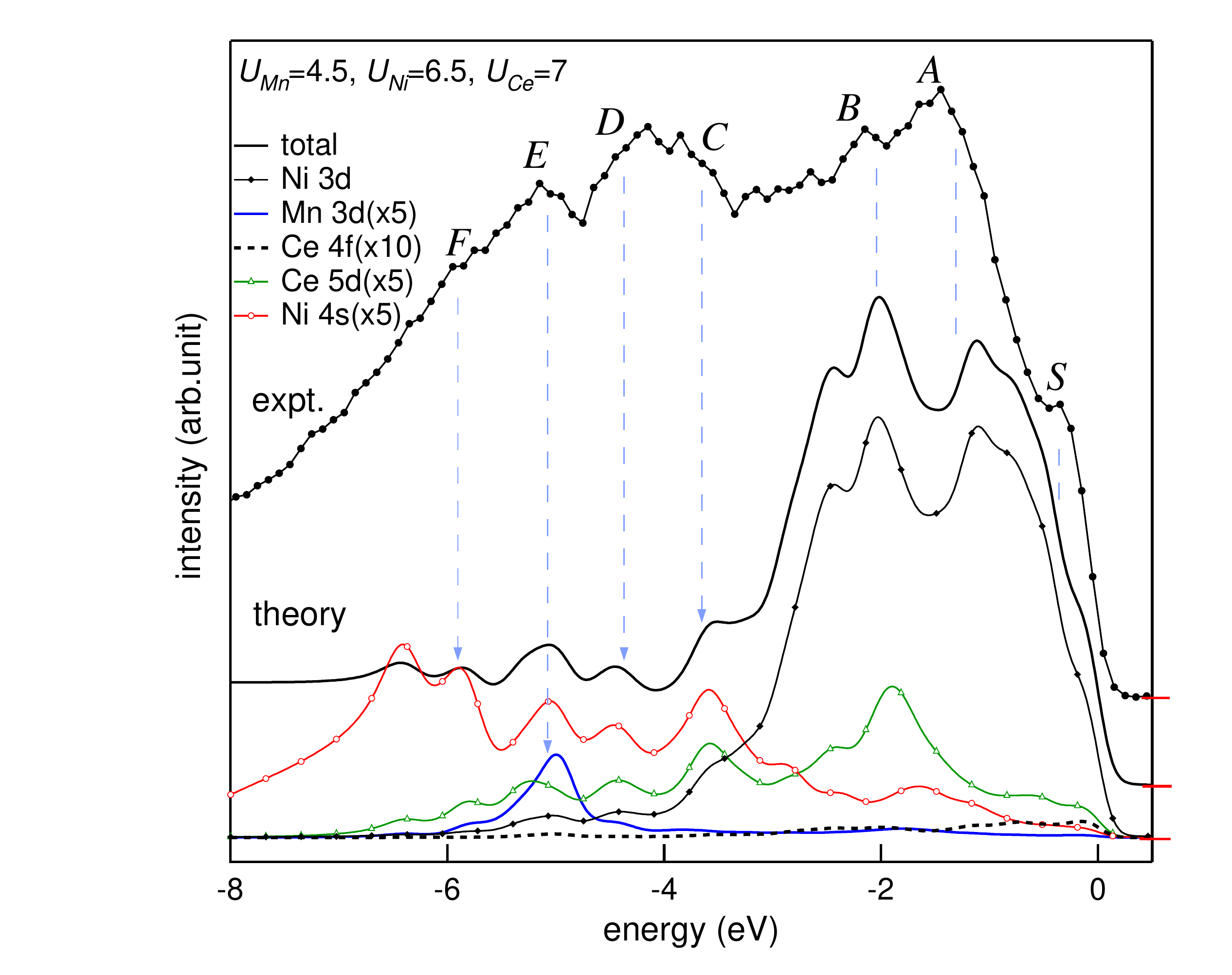} 
		\caption{ The calculated valence band spectrum for $U$(4.5,6.5,7) with the  partial  contributions  from Ni 3$d$,  Mn $3d$, Ni 4$s$,  Ce 5$d$ and Ce 4$f$ compared with experimental HAXPES VB spectrum.  The spectra are staggered along the vertical axis, and the zero for each are shown on the right vertical axis by red horizontal lines.  The peaks in the  VB are marked by $S, A, B, C, D, E$ and $F$. Light blue dashed arrows show the corresponding partial contributions to these features. $A$ and $B$ originates primarily from  Ni 3$d$ like states with small amount of contribution from Ce 5$d$ and Ni 4$s$ in $B$. Mn 3$d$ along with Ni 4$s$ like states show major contribution to $E$. $C$ and $D$ are related to Ni 4$s$, Ce 5$d$ and Ni 3$d$. Features $F$ and $S$ mainly originate from Ni 4$s$ and Ni 3$d$, respectively. The disagreement in the intensity between experiment and theory could be related to the inelastic background that is  not considered in the latter.} 
	\end{center}
	\label{S5}
	
\end{figure}
\begin{figure}[htb]
	\includegraphics[width=93mm,keepaspectratio]{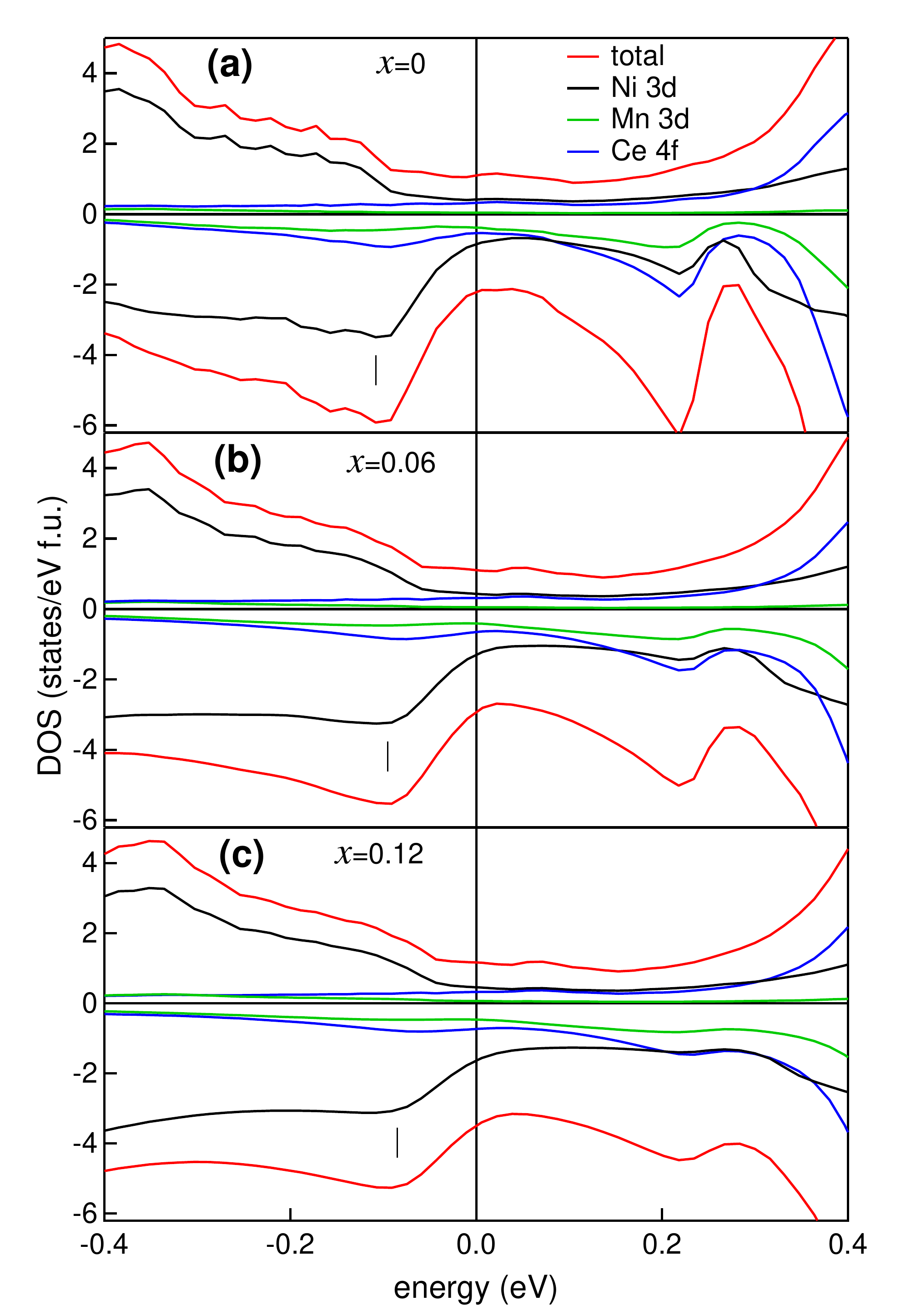} 
	
	\caption{Spin polarized total, Ni 3$d$, Mn 3$d$ and Ce 4$f$ PDOS in a small range around $E_F$,  as a function of anti-site disorder  (a) $x$= 0, (b) $x$= 0.06 and (c) $x$= 0.12.}
	\label{S6}
\end{figure}

\begin{figure}[htb]
	\includegraphics[width=93mm,keepaspectratio]{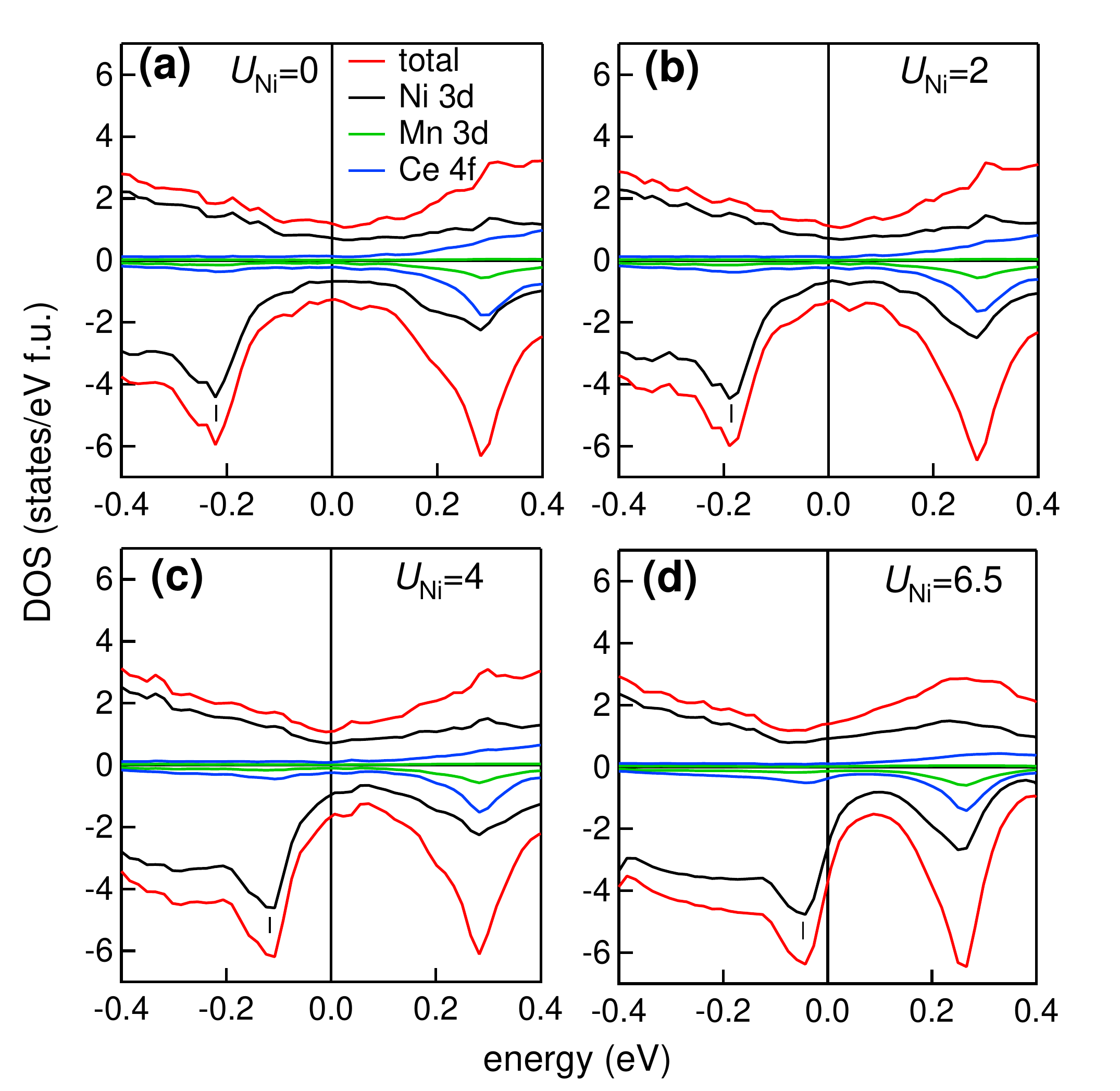} 
	\caption{Spin polarized total,  Ni 3$d$, Mn 3$d$ and Ce 4$f$ PDOS in a small range around $E_F$,  as a function of $U_{Ni}$  with $U_{Mn}$= 4.5 eV and $U_{Ce}$ =7 eV for  $U_{Ni}$= (a) 0 eV, (b) 2 eV, (c) 4 eV and (d) 6.5 eV. The black  ticks show the position of Ni 3$d$  minority spin peak that shifts towards $E_F$ with $U_{Ni}$.}
	\label{S7}
\end{figure}

\begin{table*}  
	\caption{The total spin polarization ($P_0$, also see Fig.~3 of MS) and the partial contributions to $P_0$ from the Ni 3$d$ ($P_{0_{{\rm Ni}3d}}$), Mn 3$d$ ($P_{0_{{\rm Mn}3d}}$) and Ce 4$f$ ($P_{0_{{\rm Ce}4f}}$) PDOS, as  functions of Mn-Ni anti-site disorder ($x$) and Ni 3$d$ electron-electron correlation ($U_{Ni}$). $P_0$ is calculated using the following formula:\\
		$P_0$= $|\,[n_\uparrow$($E_F$)-$n_\downarrow$($E_F$)]/[$n_\uparrow$($E_F$)+$n_\downarrow$($E_F)]\,|$,\\
		where  $n_\uparrow$($E_F$) is the majority spin total DOS at $E_F$ and 
		$n_\downarrow$($E_F$) is the minority spin total DOS at $E_F$. 
		The partial contributions to $P_0$ from an  X$nl$ PDOS  ($P_{0_{{\rm X}nl}}$) where X= Ni, Mn or Ce;  $n$=3-6; $l$= $s, p, d$ or $f$   is given by\\
		$P_{0_{{\rm X}nl}}$= $|\,[n_{\uparrow\,{\rm X}nl}$($E_F$)-$n_{\downarrow\,{\rm X}nl}$($E_F$)]/[$n_\uparrow$($E_F$)+$n_\downarrow$($E_F)]\,|$,\\ 
		where  $n_{\uparrow\,{\rm X}nl}$($E_F$) is the majority spin X$nl$  PDOS at $E_F$ and 
		$n_{\downarrow\,{\rm X}nl}$($E_F$) is the minority spin X$nl$ PDOS at $E_F$. Note that  $P_0$=  $\displaystyle\sum_{{\rm X},n,l} P_{0_{{\rm X}nl}}$, when all possible X, $n$, $l$ are considered. }
	
	\centering
	\begin{tabular}{|c|c|c|c|c||c|c|c|c|c|} \hline 
		\multicolumn{5} {|c||}  {Mn-Ni anti-site disorder} & \multicolumn{5} {|c|}  {Ni 3$d$ electron-electron correlation}    \\ \hline
		\multicolumn{1} {|c|} {$x$} 	  & \multicolumn{1} {|c|}  {$P_0$ (\%)} & \multicolumn{1} {|c|} { {$P_{0_{{\rm Ni}3d}}$}(\%) } & \multicolumn{1} {|c|}  {$P_{0_{{\rm Mn}3d}}$(\%)} & \multicolumn{1} {|c||} {$P_{0_{{\rm Ce}4f}}$ (\%) } & \multicolumn{1} {|c|} {$U_{Ni}$}  & \multicolumn{1} {|c|} {$P_0$(\%)} &  \multicolumn{1} {|c|} {$P_{0_{{\rm Ni}3d}}$(\%)} & \multicolumn{1} {|c|} { {$P_{0_{{\rm Mn}3d}}$(\%)} } & \multicolumn{1} {|c|} { {$P_{0_{{\rm Ce}4f}}$(\%)} } \\ \hline
		
		0  & 33.4  & 13.4 & 10.4 & 7 & 3 & 18.2    & 4.9 & 3.3 &  5.8\\ 
		0.04 & 40 & 18.6 & 9.2  &7.5   & 4 & 20.6    & 9.1 & 3.2 &  5.2   \\ 
		0.06 &  45 & 21.4 & 9.1  & 8.4 & 5 & 26.7 & 15 & 3 & 5.2  \\ 
		0.08  & 47 & 23.1  & 9.2 & 9.3 & 6  & 38.6   & 25.3 & 2.7 & 5.6    \\ 
		0.10& 48.4 & 24.3  & 8.8  & 8.9  & 6.5 & 45   &  32.2 & 2.4 & 5.4    \\ 
		0.12 & 50.1 & 25.3 & 8.8 & 9  & 7 & 54 & 42 & 2 & 5.2 \\ \hline
	\end{tabular}
	
	\label{pol}
\end{table*}

\section {\bf Supplementary Discussions}

\subsection {\large SD1: Discussion on the spin polarization reported in Ref. 9}
Bahramy $et~al.$\cite{Bahramy10} found the cubic phase of CeMnNi$_4$ to be stable when $U_{Mn}$ is turned on and reported that the Mn 3$d$ states shift to lower energies with $U_{Mn}$. We also find similar behavior of the Mn 3$d$ states. However, the authors also reported that the static spin polarization ($P_0$) increases substantially with $U_{Mn}$, and commented that  $U_{Ni}$ or $U_{Ce}$ have no effect on $P_0$ or any other ground state properties.  This is not in agreement with the spin polarization results we obtain here and so we discuss below the possible reason for this.

$P_0$, being proportional to the difference of  $n_\downarrow$($E_F$)  and $n_\uparrow$($E_F$),   is highly sensitive to any small change of the DOS at $E_F$.  So, it is very unlikely that  the states at $E_F$ that are dominated by Ni 3$d$ PDOS will not be influenced by $U_{Ni}$, whose value (6.5 eV) we find to be larger than $U_{Mn}$ (4.5 eV) from the comparison of the experimental HAXPES VB with DFT calculations performed by us using SPRKKR.  Our results clearly show that $U_{Ni}$ has large 
influence on both the position of the VB peaks $A$ and $B$ (Fig.~2 of MS) as well as the DOS close to $E_F$ (Fig.~3(d)). This leads to large increase of $P_0$ (Fig.~3b, Fig.~S7 and Table\,{\bf I}). 

On the other hand,  the spin polarized DOS near $E_F$ with $U_{Ni}$ was not shown in Ref.\onlinecite{Bahramy10}. Fig.~4(b,c) of that work shows the total and $s$, $d$, and $f$ PDOS for $U_{Mn}$=0 and 6 eV. Based on this figure the authors conclude that  $P_0$ increases from 10\% to 30\% with $U_{Mn}$ increasing from 0 to 6 eV.   We have analyzed this figure critically and find that the increase in $P_0$ in Ref.~\onlinecite{Bahramy10}   is due to an unusual variation of Ce 4$f$ PDOS at $E_F$ for $U_{Mn}$= 6. In their calculation, Ce 4$f$ PDOS at $E_F$ has a value of 0.3 for the minority spin, whereas in the majority spin PDOS it is largely reduced to 0.02. Thus, 
increase in $P_0$ to 30\% for  $U_{Mn}$= 6 eV  results primarily from the variation of the Ce 4$f$ spin polarized states. Thus, we find  $P_{0_{{\rm Ce}4f}}$ (defined in Table\,{\bf I} of SM) increases from 7.9\%  to 30\%, a whopping 380\% increase, while $P_0$ increases from 10 to 30\%. Thus at $U_{Mn}$= 6 eV, the whole spin polarization is contributed by Ce 4$f$ states only. 

The above discussed effect of $U_{Mn}$ on $P_0$  and Ce 4$f$ states  is {\bf unlikely}, 
since the Mn 
atom is surrounded by  the Ni atoms and Ce is only the second nearest neighbor at a large distance of 3.02\AA. Rather, one would expect that the Ni 3$d$ states that are dominant at $E_F$ would  contribute to $P_0$, but between Fig.~4(b) and Fig.~4(c) of Ref.~\onlinecite{Bahramy10}, the 3$d$ states hardly change (for $U_{Mn}$=0 these are 0.3  and 0.5 states/eV for majority and minority spin, respectively and  for  $U_{Mn}$= 6 eV these are 0.2 and 0.4 states/eV for majority and minority spin, respectively). Thus,  strangely, $P_{0_{3d}}$  $i.e.$ spin polarization due to the 3$d$ states remains essentially similar ($P_{0_{3d}}$= 13\% for $U_{Mn}$= 0 and $P_{0_{3d}}$= 16\%  $U_{Mn}$= 6 eV). Thus, while $P_{0_{{\rm Ce}4f}}$  increases by 380\%, $P_{0_{3d}}$ hardly changes.  This  seems to be an unphysical result.

We find that the increase of $P_0$ for both anti-site disorder ($x$) as well as electron-electron correlation of the Ni 3$d$ electrons ($U_{Ni}$) is primarily due to the changes in the Ni 3$d$ PDOS. Table\,{\bf I} of SM clearly shows how only      $P_{0_{{\rm Ni}3d}}$ increases as $P_0$, while $P_{0_{{\rm Ce}4f}}$ and $P_{0_{{\rm Mn}3d}}$ remain almost unchanged and thus do not play any role in the enhancement of $P_0$.
\begin{figure}[htb]
	\begin{center}
		\includegraphics[width=93mm,keepaspectratio]{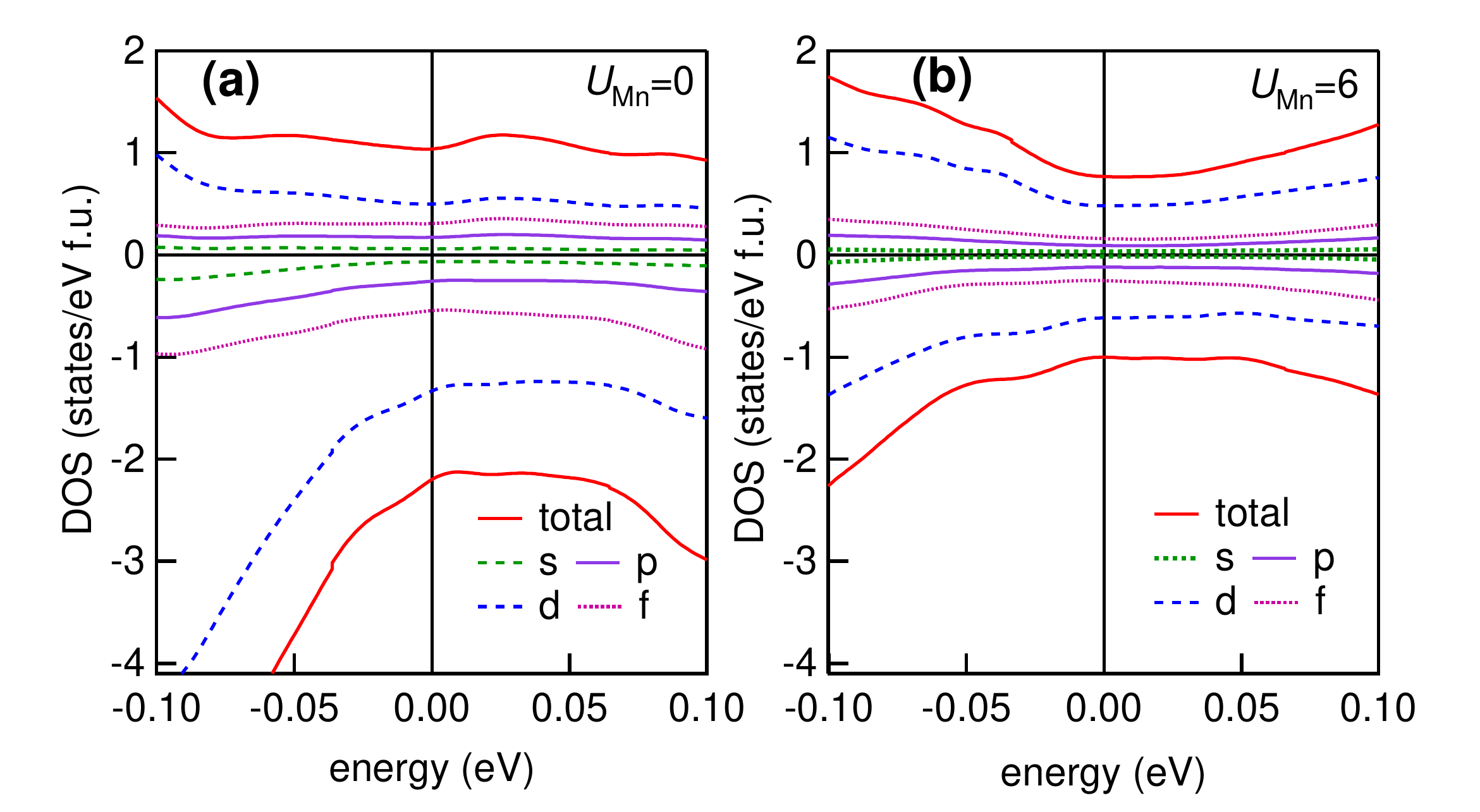} 
		\caption{Spin polarized total DOS and total $s$, $p$, $d$ and $f$ PDOS  of CeMnNi$_{4}$   with (a) $U_{Mn}$=0 eV (b) $U_{Mn}$= 6 eV for comparison with Fig.~4(b,c) of Ref.~\onlinecite{Bahramy10}.}
		\label{SPdoswithUMn}
	\end{center}
	\label{S8}
\end{figure}
For comparison with Ref.~\onlinecite{Bahramy10},  in Fig.~S8, we show the total and PDOS  with $U_{Mn}$=6 eV, with $U_{Ni}$= $U_{Ce}$=0.  We find that   $P_0$ decreases from 33.4\% to 13.2\% with $U_{Mn}$ varying from 0 to 6 eV (Fig.~4(a) in MS).
It is evident that  the total $f$ PDOS  is  less than the total $d$ PDOS over the entire range  and most importantly remains  similar between $U_{Mn}$=0 and $U_{Mn}$= 6 eV. Thus,  $P_{0_{4f}}$   is almost unchanged (rather decreases slightly)  from 7.3\% to 5.3\%  from $U_{Mn}$= 0 to 6 eV. This is in stark disagreement with the 380\% increase of $P_{0_{4f}}$ that can be concluded from Ref.~\cite{Bahramy10}.

\subsection {\large SD2: Surface valence transition}

\begin{figure}[htb]
	\centering
	\includegraphics[width=93mm,keepaspectratio]{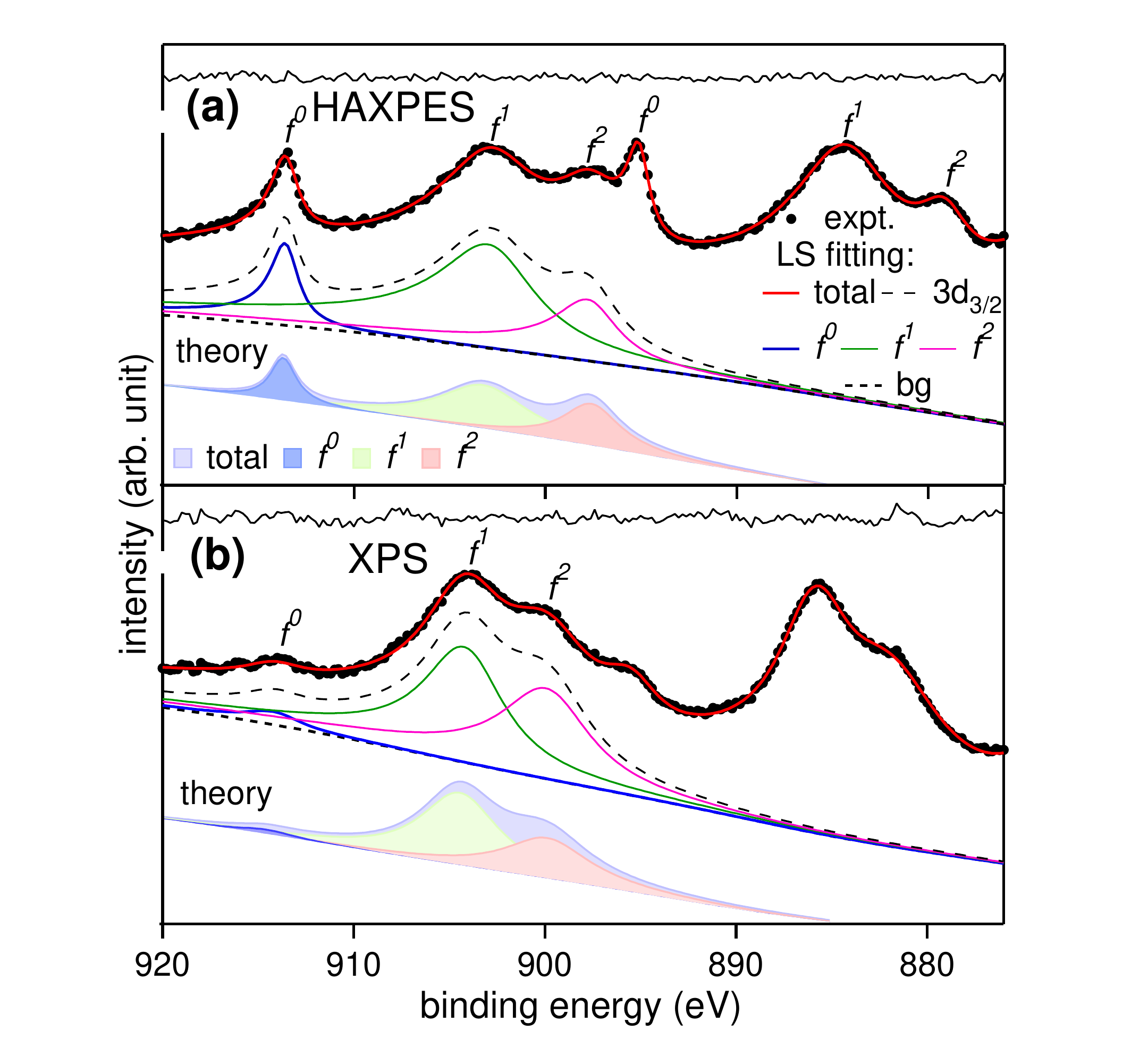}
	\caption{ Ce 3$d$  core-level spectra (black dots) recorded with (a) 8 keV (HAXPES) and (b) 1.48 keV (XPS) photon energies. The spectra have been fitted (red line) using a least square (LS) error minimization routine  and the $f^{n}$ satellite components for Ce 3$d_{3/2}$ are shown.
		~ The calculated Ce 3$d_{3/2}$ spectra using IW theory along with the $f^{n}$ satellites are shown at the bottom and  the residuals of fitting (black line) are shown at the top of each panel.   }
	\label{Ce3d}
\end{figure}

The Ce 3$d$  core-level spectrum  displays two sets of triplet peaks corresponding to the spin-orbit split components (Fig.~S\ref{Ce3d}). The most intense among the triplet peaks is the $f^1$ satellite associated with a poorly screened 3$d^{9}$4$f^{1}$ final state occurring  at   902.8 eV and 884.4 eV  binding energies. The two additional satellite peaks that occur at relatively higher and lower binding energies are referred to as  $f^{0}$ and $f^{2}$, respectively.  The well screened $f^{2}$ satellite has  an extra screening electron with 3$d^{9}$4$f^{2}$ final state, while the $f^{0}$ satellite is related to 3$d^{9}$4$f^{0}$ final state\cite{Hillebrecht82,Fuggle83}. Notable in Fig.~S\ref{Ce3d} is the large $f^0$
intensity in HAXPES, which decreases drastically in  soft x-ray PES (XPS). In order to extract quantitative information, 
the Ce 3$d$ core-level spectra were fitted using a least square error minimization routine with each peak assigned a  Doniach and  $\breve{S}$unji$\acute{c}$ (DS) line shape\cite{Doniach70}. This was further convoluted with a Gaussian function of fixed width to represent the instrumental broadening. 
~Since Ni 2$p$ that appears close in binding energy to Ce 3$d$  might contribute to the intensity in the Ce 3$d$ region, the Ni 2$p$ main and satellite peaks were also included in the fitting scheme. The whole region including Ni 2$p$ along with the components is shown in Fig.~S10. A total 10 DS line shapes were used, 6 for Ce 3$d$ comprising of the three $f^n$ components for each spin-orbit (s.o.) peaks and 4 for  Ni 2$p$ representing the main peak and  satellite for both the s.o. components. The  parameters defining each  DS line shape are the  intensity, position, width ($\Gamma$) and asymmetry parameter ($\alpha$). A Touguaard background was also included in the fitting scheme, where the $B_1$ parameter was varied and the $C$ parameter was kept fixed at 1643 eV$^2$\cite{Tougaard89}. Thus, a total of 35 parameters defined the full spectral shape including Ce 3$d$ and Ni 2$p$. However, some reasonable constraints were needed, for example (i) the life time broadening of $f^0$ for Ce 3$d_{3/2}$ was constrained to be greater than or equal to $f^0$ for Ce 3$d_{5/2}$, (ii) $\alpha$  was kept equal for all Ce 3$d$ DS components, (iii) for XPS fitting, the satellites of Ni 2$p$ have same width as HAXPES.

\begin{figure}[htb]
		\includegraphics[width=93mm,keepaspectratio]{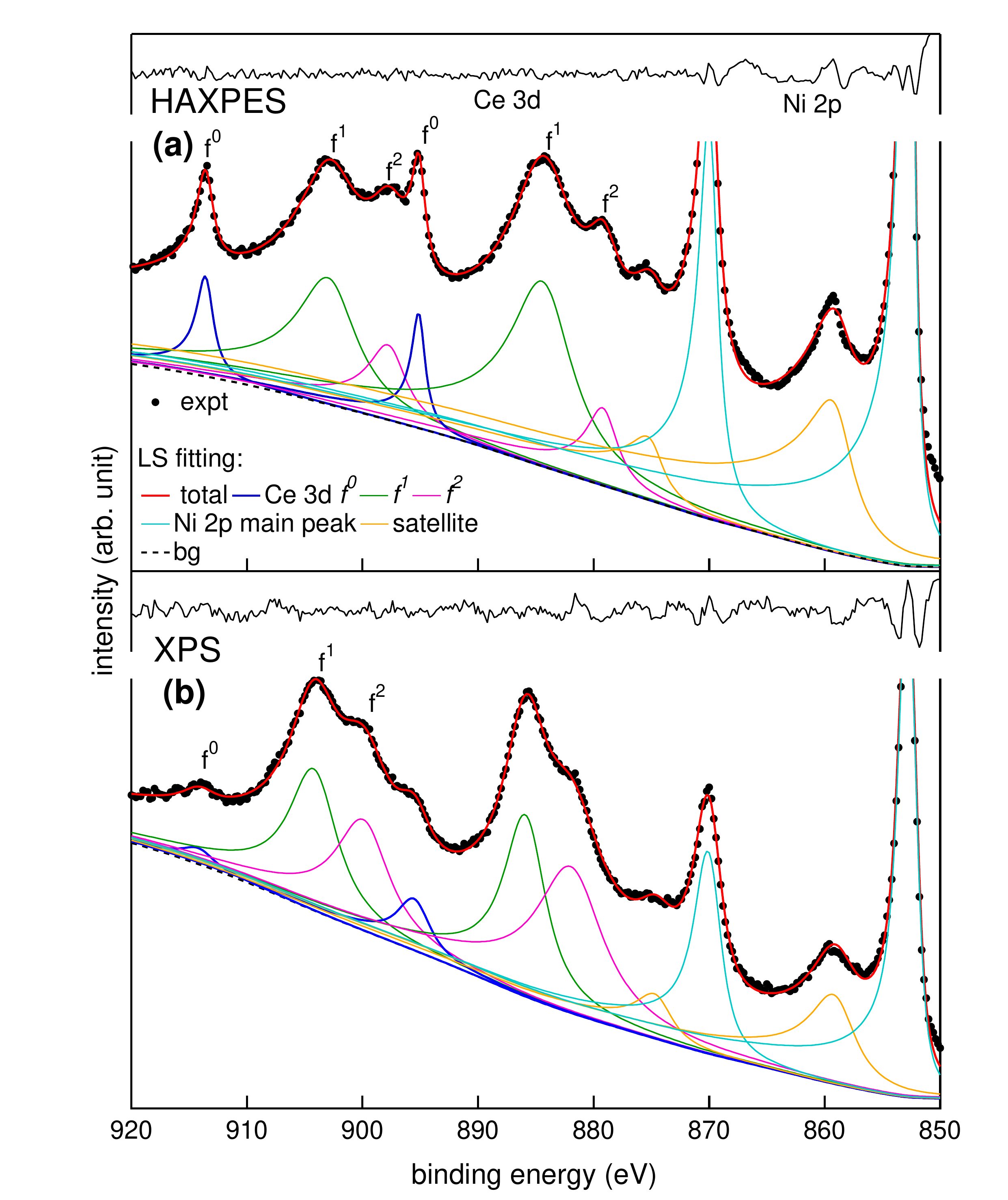} 
		\caption{Ce 3$d$ core-level spectra (black circles, includes Ni 2$p$ region)  with (a)  HAXPES compared with (b) XPS. The spectra has been fitted (red line) using a least square (LS) error minimization routine, and the residuals (black line) are shown at the top of each panel. The different components such as  Ce 3$d$ $f^{0}$, $f^{1}$, and  $f^{2}$, Ni 2$p$ main peaks and  satellites as well as a Tougaard background are shown.}
		\label{S10}
	\label{11}
\end{figure}


From the least square fitting, we find that  the normalized intensity of  $f^0$ ($I_n(f^0)$)  is  0.15 for HAXPES, where $I_n(f^0)$= $I(f^0)$/$\displaystyle\sum_{n=0}^2 I(f^n) $) (Table\,{\bf II} of SM). Such large intensity of  $f^0$ having almost similar height as $f^1$ is unusual and has not been observed in other Ce based intermetallic compounds\cite{Yano08,Sundermann16}. In contrast, $I_n(f^0)$  is an order of magnitude less (0.04) in XPS. 
~This could be related to the bulk sensitivity of HAXPES with  mean free path ($\lambda$) of  91 \AA~ for Ce 3$d$ electrons
~while  XPS is surface sensitive with  $\lambda$= 13 \AA\cite{tpp2m}. 
~In order to  understand the differences between the above discussed bulk and surface Ce 3$d$
spectra, we turn to a simplified version of the  Anderson single-impurity model\cite{Gunnarsson83} proposed by Imer and Wuilloud (IW), where the extended valence states are considered as a band of infinitely narrow width\cite{Imer87}. The Ce 3$d$ spectrum is calculated as a function of the energy of the unhybridized $4f$ state relative to $E_F$ ($\epsilon_f$),   Coulomb repulsion between 4$f$ electrons at the same site ($U_{ff}$),  Coulomb attraction between  4$f$ electron and the final-state core hole ($U_{fc}$), and hybridization  between the 4$f$ states and the conduction band ($\Delta$).

\begin{table*}  
	\caption{The parameters for Ce 3$d_{3/2}$ such as normalized intensity $I_n$($f^n$), 
		binding energy ($E_B$), energy separation $\delta_{0n}$ between $f^n$ satellites obtained from least square fitting  compared with those obtained from IW theory. All the values are in eV and $U_{ff}$= 7 eV in both cases. The error in  $I_n$($f^0$)  is significantly smaller because the $f^0$ satellite it is well separated in energy from all the other components.} 
	
	\centering
	\begin{tabular}{|c|c|c|c|c||c|c|c|c|c|} \hline 
		\multicolumn{1} {|c|}  {h$\nu$  } & \multicolumn{4} {|c||}  {PES experiment} & \multicolumn{5} {|c|}  {IW theory}    \\ \hline
		\multicolumn{1} {|c|} {(keV)} 	  & \multicolumn{1} {|c|}  {$f^n$} & \multicolumn{1} {|c|} { {$I_n$} } & \multicolumn{1} {|c|}  {$E_B$$\pm$0.2} & \multicolumn{1} {|c||} {$\delta_{0n}$$\pm$0.4  } & \multicolumn{1} {|c|} {$\epsilon_f$}  & \multicolumn{1} {|c|} {$\Delta$} &  \multicolumn{1} {|c|} {$U_{fc}$} & \multicolumn{1} {|c|} { {$I_n$} } & \multicolumn{1} {|c|} { {$\delta_{0n}$} } \\ \hline
		
		8  & $f^{0}$ & {\bf 0.15$\pm$0.01} & 913.5 & 0 &  & &  & {\bf 0.15} & 0  \\ 
		& $f^{1}$ & 0.6$\pm$0.1 & 902.8  & 10.7  & -1.0 & 1.5    & 10 & 0.53 &  10.5   \\ 
		& $f^{2}$ & 0.25$\pm$0.1 &  897.8  & 15.7 &  & & & 0.32 & 16.1  \\ \hline

		1.48  & $f^{0}$ & {\bf 0.04$\pm$0.01}  & 914.3 & 0 &   &   & & {\bf0.04} & 0    \\ 
		& $f^{1}$ & 0.5$\pm$0.1 & 904.1 &  10.2 & -2.5 & 1.1   &  8 & 0.62 & 10.1    \\ 
		& $f^{2}$ & 0.45$\pm$0.1 & 899.9 & 14.4  & & & & 0.34 & 14.7 \\ \hline
	\end{tabular}
	\label{Ce3dtable}
\end{table*}

The above mentioned parameters are varied such that the $f^n$ satellites of the calculated Ce 3$d_{3/2}$ spectrum have similar intensities ($I_n$) and energy separations between $f^0$ and $f^n$ ($\delta_{0n}$), as obtained from the fitting of the experimental spectra.    For example, besides the large change in $I_n$($f^0$), the binding energies  of the $f^n$ satellites are lower  in HAXPES (this is not due to recoil effect\cite{Fadley10}, see Fig.~S11), resulting in different $\delta_{0n}$ as shown in Table\,{\bf II}.
\begin{figure}[htb]
	\includegraphics[width=93mm,keepaspectratio]{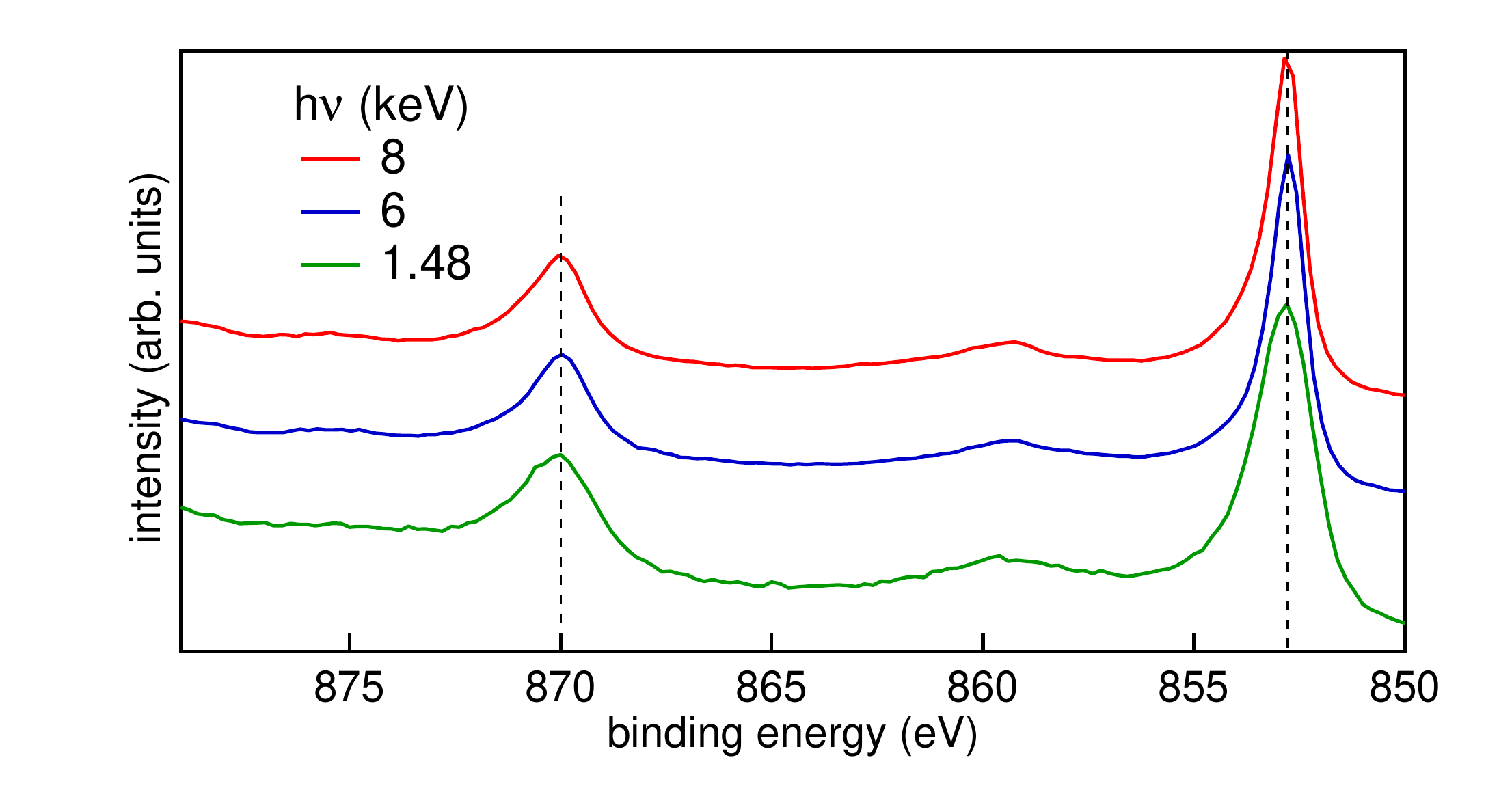} 
	\caption{Ni 2$p$ core level spectra of CeMnNi$_4$ using 8, 6   and 1.48 kV photon energies, normalized to same height at the  Ni 2$p_{3/2}$ peak and staggered along the vertical axis for clarity of presentation.
		The recoil effect that has been observed in the HAXPES spectra of light materials\cite{Fadley10} is absent here since it comprises of heavier 3$d$ and rare earth elements. The recoil effect, if present causes a uniform shift of the  peaks to higher binding energies that increases with the kinetic energy of the electrons, which in turn depends on the photon energy used.  We confirm the absence of any recoil effect here  from the  Ni 2$p$ spectra taken with different photon energies where any shift of the peaks for different photon energies is absent.}
	\label{S11}
\end{figure}

\begin{figure}[htb]
	\includegraphics[width=93mm,keepaspectratio]{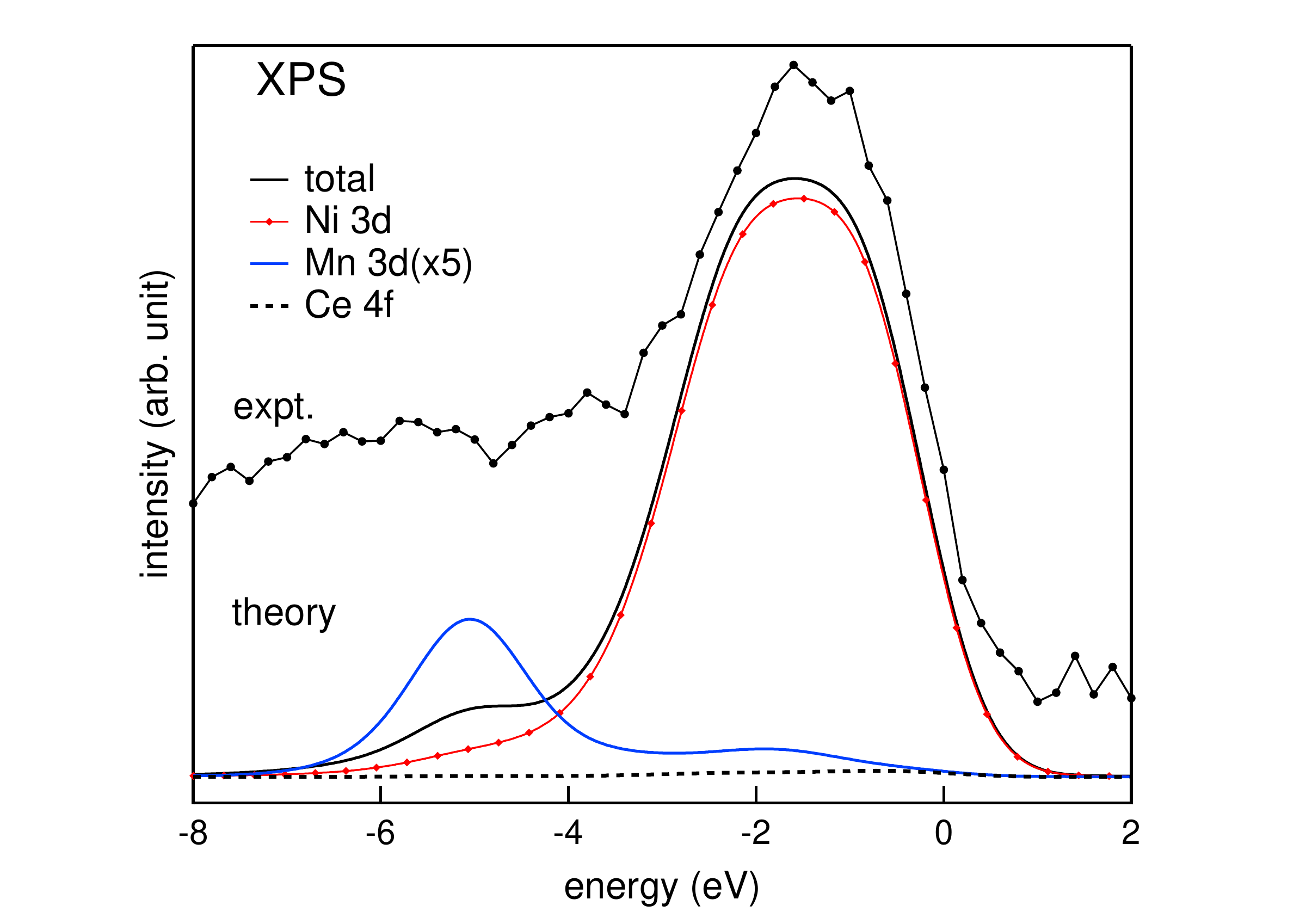} 
	\caption{The XPS valence band spectrum of CeMnNi$_4$ taken with 1.25 keV photon energy (black line with dots) compared with calculated VB along with the Ni 3$d$, Mn 3$d$ ($\times$5) and Ce 4$f$ partial contributions. }
	\label{S12}
\end{figure}
\begin{figure}[htb]
	\includegraphics[width=90mm,keepaspectratio]{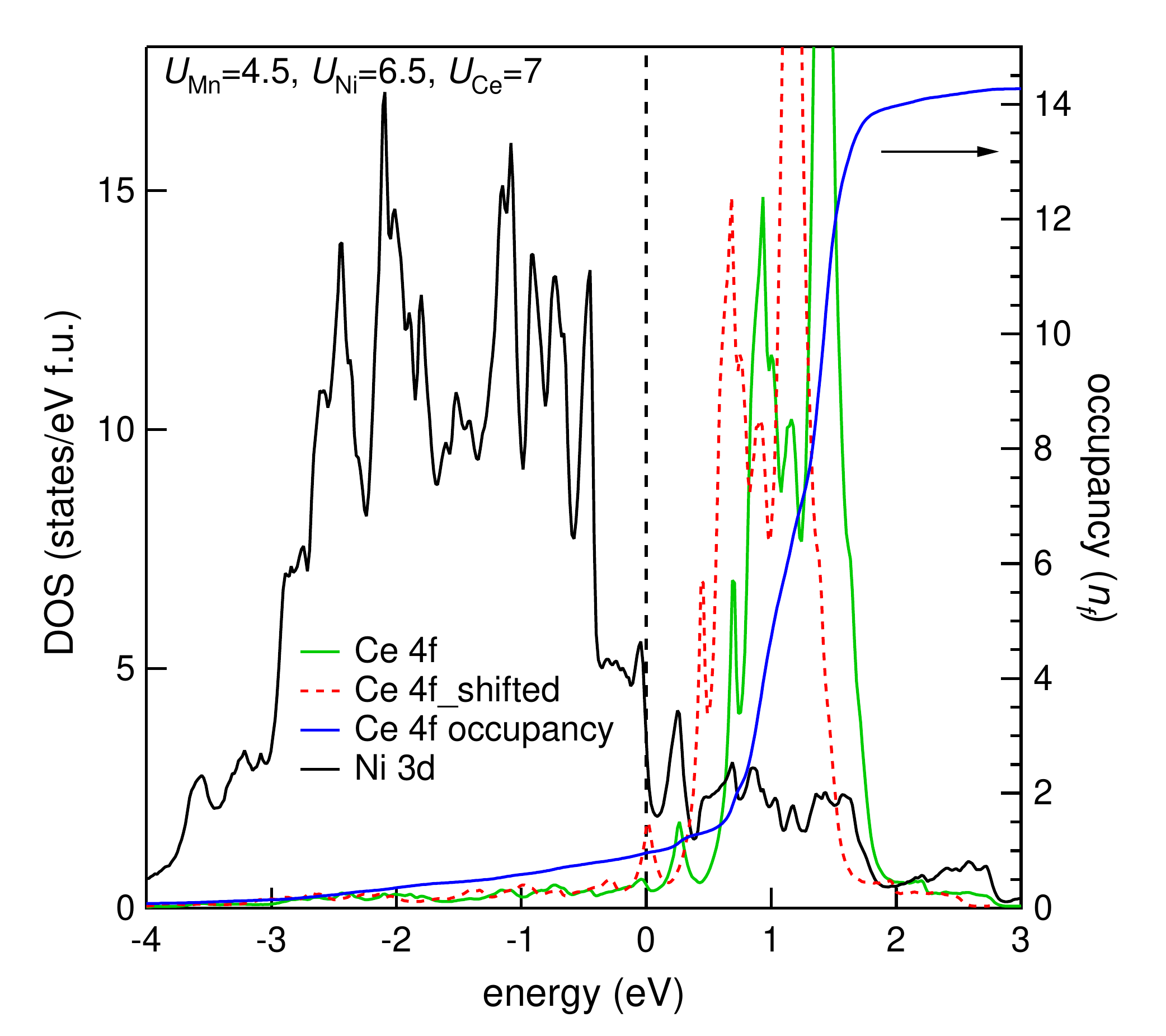} 
	\caption{Spin integrated Ni 3$d$ and Ce 4$f$ PDOS for CeMnNi$_4$ with optimum $U(4.5,6.5,7)$. The integration of the Ce 4$f$ PDOS (blue line) that shows the occupancy ($n_f$) on the right axis.}
	\label{S13}
\end{figure}

~
~In order to simulate the Ce 3$d$ HAXPES spectra using IW theory, we note that $I_n$($f^0$)  increases sensitively with $\epsilon_f$, and so this parameter is varied keeping the others fixed at the values  suggested for Ce compounds ($\Delta$= 1.5 eV, $U_{ff}$= 7 eV, $U_{fc}$= 10 eV)\cite{Imer87}.  For $\epsilon_f$= -1 eV, we find $I_n$($f^0$)= 0.15  in excellent agreement with experiment; and the other quantities such as $\delta_{0n}$,  $I_n$($f^1$) and $I_n$($f^2$) are also in good agreement (Table~{\bf II}). The calculated spectrum obtained with $\epsilon_f$= -1 eV, $\Delta$= 1.5 eV, $U_{ff}$= 7 eV, $U_{fc}$= 10 eV   is shown at the bottom of Fig.~S\ref{Ce3d}(a), where the  $f^n$ satellites have been broadened by their respective widths obtained from the fitting  and a  background\cite{Tougaard89} has also been added.    The $f$ occupancy in the  ground state ($n_f$) turns out to be 0.8, indicating a mixed valent state with 20\% Ce in $f^0$ (Ce$^{4+}$) while 80\% in $f^1$ (Ce$^{3+}$) configuration.


In order to simulate the Ce 3$d$ XPS spectrum,   we decrease  $\epsilon_f$  to -2.5 eV from the HAXPES value of -1 eV and  obtain $I_n$($f^0$)= 0.04. But  concomitantly, both $\delta_{01}$ (=12.1 eV) and $\delta_{02}$ (=18.9 eV) become  larger than experimental values of 10.2 eV and 14.4 eV, respectively (Table\,{\bf II}). In order to decrease  $\delta_{0n}$, both $\Delta$ and $U_{fc}$ need to be decreased, and thus, we obtain a good agreement with experiment for   $\epsilon_f$= -2.5 eV, $\Delta$= 1.1 eV, $U_{fc}$= 8 eV, and $U_{ff}$= 7 eV 
~(bottom of Fig.~S\ref{Ce3d}(b)).  
Due to the decrease of $\epsilon_f$, $n_f$ increases to 0.98, and thus,  in contrast to bulk, at the surface Ce has predominantly  3$d^{9}$4$f^{1}$ (Ce$^{3+}$) ground state. Thus, in the bulk, since $\epsilon_f$ (= -1 eV) is closer to $E_F$ and $\Delta$ is larger, the Ce 4$f$ electron transfers to the valence states comprising of primarily Ni 3$d$ states making CeMnNi$_4$ a  mixed valent system with 4$f$ occupancy  of $n_f$= 0.8. However, at the surface, the reduced hybridization between the  Ce 4$f$ and unsaturated 3$d$ states results in a lowering of the Ce 4$f$ states further below $E_F$. This increases the occupancy of the Ce 4$f$ level ($n_f$= 0.98) and results in the surface valence transition. Decrease in $U_{fc}$ from about 10 eV to 8 eV at the surface is also a manifestation of this transition  possibly caused by the more efficient screening of the core hole due to increased $n_f$.  


It might be noted that although the  surface valence transition is clearly  manifested in the Ce 3$d$ core-level spectra,   it does not however result in appearance of any Ce 4$f$ peak in the XPS VB (Fig.~S12), 
~ which could be expected due to enhanced $n_f$ at the surface. Firstly, this happens because  the occupied part of  Ce 4$f$ PDOS  from -3 eV to $E_F$ is largely diminished, broad and featureless  (Fig.~S13). 
~Its integration (blue line) up to $E_F$  gives  $n_f$= 0.96  in the bulk from DFT, which is in reasonable agreement with $n_f$= 0.8 from IW method, considering  the  assumptions of  the latter model calculation\cite{Imer87}.  The increase of $n_f$  by 0.18  at the surface obtained from IW method would manifest itself  through a small shift of the Ce 4$f$ PDOS by 0.25 eV (obtained from integration of PDOS that gives $n_f$= 1.14) to lower energy in the rigid band model (red dashed line in Fig.~S13). Thus, the main peak of Ce 4$f$ still remains above $E_F$ at the surface.  Secondly,  occupied Ni 3$d$ PDOS  as well as its photoemission cross-section\cite{Yeh85} are  much larger than Ce 4$f$ (Fig.~S12, S2-S4) resulting in complete domination of the VB by Ni 3$d$ states at low photon energies too. This is reconfirmed by the relative contributions of  Ni 3$d$ and  Ce 4$f$  to the calculated XPS  VB in Fig.~S12. 
\end{document}